\documentclass[twocolumn,showpacs,preprintnumbers,amsmath,amssymb,prb,superscriptaddress]{revtex4-1}
\usepackage{bbm}
\usepackage{mathrsfs}
\usepackage{graphicx}
\usepackage{dcolumn}
\usepackage{bm}
\usepackage{amsmath}
\usepackage{amsfonts}
\usepackage{color}
\usepackage{floatrow}

\begin{document}

\title{Electrically tunable topological superconductivity and Majorana fermions in two dimensions}
\author{Jing Wang}
\affiliation{State Key Laboratory of Surface Physics and Department of Physics, Fudan University, Shanghai 200433, China}
\affiliation{Collaborative Innovation Center of Advanced Microstructures, Nanjing 210093, China}

\begin{abstract}
The external controllability of topological superconductors and Majorana fermions would be important both for fundamental and practical interests. Here we predict the electric-field control of Majorana fermions in two-dimensional topological superconductors utilizing a topological insulator thin film proximity coupled to a conventional $s$-wave superconductor. With ferromagnetic ordering, the tunable structure inversion asymmetry by vertical electric field could induce topological quantum phase transition and realize a chiral topological superconductor state. A zero-energy Majorana bound state appears at the boundary of an applied electric field spot, which can be observed by scanning tunneling microscopy. Furthermore, the structure inversion asymmetry could also enlarge the helical topological superconductor state in the phase diagram, making the realization of such an exotic state more feasible. The electrical control of topological phases could further apply to van der Waals materials such as two-dimensional transition metal dichalcogenides.
\end{abstract}

\date{\today}


\maketitle

\section{Introduction}

Topological superconductors (TSCs) are new states of quantum matter which are characterized by a fully paring gap in the bulk, and topologically protected gapless edge states consisting of Majorana fermions~\cite{hasan2010,qi2011,tanaka2012,beenakker2016,sato2016}. Majorana fermion, being its own antiparticle, has potential applications in future quantum computations~\cite{nayak2008,alicea2012,beenakker2013}. There are two basic types of TSC states in two dimensions (2D). The first is the time-reversal breaking (TRB) chiral TSC characterized by Chern number $\mathcal{N}$ in class D~\cite{schnyder2008}, which has $\mathcal{N}$ gapless chiral Majorana edge fermions. Moreover, a Majorana zero mode is predicted to be trapped in each vortex core~\cite{read2000}, which leads to non-Abelian statistics of the vortex~\cite{ivanov2001}. The second is the time-reversal invariant (TRI) helical TSC with a $Z_2$ topological index in class DIII. It has an odd number of counterpropagating helical Majorana fermion edge states protected by the TR symmetry, and an emergent supersymmetry is naturally present in this system~\cite{qi2009}. There are tremendous efforts to search for both the chiral~\cite{mackenzie2003,fu2009a,sato2009,sau2010,alicea2010,qi2010b,chung2011,wang2015c,ojanen2015} and helical~\cite{tanaka2009,sato2009a,nakosai2012,wang2014c} TSCs in 2D, however, the experimental situation is still not definitive.

A promising candidate for chiral TSC is to utilize the quantum anomalous Hall (QAH) insulator~\cite{chang2013b,wang2015d} proximity coupled to a conventional $s$-wave superconductor~\cite{qi2010b,wang2015c,chung2011}. The key idea is that in this hybrid system, a chiral TSC phase with odd number of chiral Majorana edge fermions always emerges in the neighborhood of the transition between the QAH and normal insulator (NI)~\cite{qi2010b}, which can be driven by an external magnetic field~\cite{wang2015c}. Recently, such a hybrid structure using magnetic topological insulators (TIs) has been fabricated experimentally~\cite{he2016}. A series of topological states are controlled by the magnetization reversal, where a narrow half-integer quantized conductance plateau is observed at the coercive field as a compelling evidence of the \emph{single} chiral Majorana edge state~\cite{wang2015c,he2016,lian2016,law2016}. To further confirm such a state is a chiral TSC with an odd integer Chern number $\mathcal{N}$=1, one need to demonstrate the zero-energy Majorana bound state (MBS) in the vortex core and prove its non-Abelian nature. However, such a chiral TSC state resides at the coercivity, where multiple random magnetic domains form in the parent magnetic TIs~\cite{wang2014a,liu2016}. Therefore the system is inhomogeneous, which makes the observation of MBSs much more difficult. Here we propose a new route to control the topological quantum phase transition (QPT) and realize the chiral and helical TSCs, which are generic in layered materials coupled to $s$-wave superconductors. Take magnetic TIs as a model example, we find that the structure inversion asymmetry (SIA) by applying an electric field could induce topological QPT and realize a $\mathcal{N}$=1 chiral TSC. Interestingly, a zero-energy MBS appears at the boundary of an applied electric field spot, which can be observed by scanning tunneling microscopy. Compared to the magnetic control~\cite{wang2015c,he2016}, the electric manipulation of the chiral TSC proposed here has two advantages. First, the parent magnetic TI under the external electric field should be homogenous. Second, the superconducting proximity effect should be much stronger without external magnetic field. Furthermore, in the TRI case, we find SIA could enlarge the helical TSC state in the phase diagram, making the realization of such an exotic state more feasible.

The organization of this paper is as follows. After this introductory section, Sec.~\ref{model} describes the effective model for the QAH and QSH state in a TI thin film with and without FM ordering, respectively, where the topological properties tuned by SIA has been studied.
Section~\ref{chiral} presents the results on the chiral TSC state, the phase diagram, edge state, and estimation of SIA in realistic materials. Section~\ref{MBS} is devoted to the prediction and experimental feasibility of the zero-energy MBS created by an applied electric field spot in a magnetic TI.
Section~\ref{helical} presents the results on the TRI helical TSC state. Section~\ref{discussion} presents discussion on the possible experimental realization of TSC in other van der Waals materials. Section~\ref{conclusion} concludes this paper. Some auxiliary materials are relegated to Appendixes.

\section{Model}
\label{model}

The physical effects discussed in this paper is generic for any layered QAH and QSH insulator materials. We would like to start with studying the topological properties of a 2D TI thin film with and without ferromagnetic (FM) order. The low-energy physics of the system consists of the Dirac-type surface states only~\cite{wang2014a}. The 2D effective Hamiltonian is $\mathcal{H}_0(\vec{k})=\sum_{\vec{k}}\psi^{\dag}_{\vec{k}}H_0(\vec{k})\psi_{\vec{k}}$, with
\begin{eqnarray}\label{model0}
H_0(\vec{k}) &=& v_Fk_y\sigma_1\otimes\tau_3-v_Fk_x\sigma_2\otimes\tau_3+m(\vec{k})1\otimes\tau_1
\nonumber
\\
&&+\Lambda\sigma_3\otimes1+V(x,y)1\otimes\tau_3,
\end{eqnarray}
and the field operator $\psi_{\vec{k}}=(c_{t\uparrow},c_{t\downarrow},c_{b\uparrow},c_{b\downarrow})^T$, where $t$ and $b$ denote the top and bottom surface states and $\uparrow$ and $\downarrow$ represent spin-up and spin-down states, respectively. $\vec{k}=(k_x,k_y)$. $v_F$ is the Fermi velocity, where $\hbar$ is absorbed into the definition of $v_F$. $\sigma_i$ and $\tau_i$ ($i=1,2,3$) are Pauli matrices acting on spin and layer, respectively. $m(\vec{k})=m_0+m_1(k_x^2+k_y^2)$ describes the tunneling effect between the top and bottom surface states, and set $m_1>0$. $\Lambda$ is the mean value of exchange field along the $z$ axis introduced by the FM ordering. In magnetically doped TIs such as Cr~\cite{chang2013b} and V~\cite{chang2015}, $\Lambda\neq0$. For TRI system without magnetic dopants, $\Lambda=0$. $V$ denotes the SIA between the two layers, which may have spatial dependence, and can be tuned locally or globally by applying an electric field along the $z$ direction.

The topological properties of the system depend on the SIA, which is clearly seen by a basis transformation into the following form
\begin{equation}
\begin{aligned}
&\widetilde{H}_0(\vec{k}) =
\begin{pmatrix}
\widetilde{H}_+(\vec{k}) & V\sigma_1\\
V\sigma_1 & \widetilde{H}_-(\vec{k})
\end{pmatrix},
\\
\widetilde{H}_\pm(\vec{k}) &= v_F(k_y\sigma_1\mp k_x\sigma_2)+(m(\vec{k})\pm\Lambda)\sigma_3.
\end{aligned}
\end{equation}
Here the basis for $\widetilde{H}_0(\vec{k})$ is $(c_{t\uparrow}+c_{b\uparrow},c_{t\downarrow}-c_{b\downarrow},c_{t\downarrow}+c_{b\downarrow},c_{t\uparrow}-c_{b\uparrow})^T/\sqrt{2}$. First we consider $\Lambda\neq0$ with TRB. If $V=0$, the system is decoupled into two models, and the Chern number of $\widetilde{H}_\pm(\vec{k})$ depends only on the sign of Dirac mass $m(\vec{k})\pm\Lambda$ at $\vec{k}=0$. When $|\Lambda|>|m_0|$, the total Chern number is $C=\Lambda/|\Lambda|$ and the system is a QAH insulator. When $|\Lambda|<|m_0|$, $C=0$ and the system is a NI. As $\Lambda$ decreases during the flipping of the magnetic domains at the coercivity, a topological QPT from QAH to NI happens~\cite{kou2015,fengy2015}. If uniform $V\neq0$, such SIA term may also induce topological QPT from a QAH to NI. The phase boundary [Fig.~\ref{fig1}(a)] is determined by the bulk gap closing in the spectrum as $m_0^2+V^2=\Lambda^2$, with critical point $V^c_a=\sqrt{\Lambda^2-m_0^2}$. Here $V^c_a$ is the critical value of $V$ for the topological QPT from QAH to NI. For $V<V_a^c$, the system is adiabatically connected to the QAH state with a full gap. For $V>V_a^c$, the system is a NI. Second, in the TRI case with $\Lambda=0$, the topological properties in the presence of SIA term was discussed in Ref.~\cite{shan2010,wang2015a}. For $V<V_s^c$, the system is a QSH insulator. For $V>V^c_s$, the system is a NI. Here $V_s^c=v_F\sqrt{-m_0/m_1}$ is critical value of $V$ for the topological QPT from QSH to NI.

\section{Chiral TSC}
\label{chiral}

Physically, a QAH state with Chern number \emph{C}=1 in proximity with an $s$-wave superconductor is naturally a chiral TSC with $\mathcal{N}$=2 chiral Majorana edge states. Therefore, a $\mathcal{N}$=1 chiral TSC phase emerges in the neighborhood of the transition between the QAH and NI~\cite{qi2010b,wang2015c}. As discussed in Sec.~\ref{model} clearly, the topological QPT from QAH to NI could be driven by SIA. So a $\mathcal{N}$=1 chiral TSC induced by SIA is expected. To confirm the validity of the picture, we first analyze the bulk Chern number of the chiral TSC as a function of SIA; and then calculate the Majorana edge spectrum in the effective model. The Bogoliubov-de Gennes (BdG) Hamiltonian for the $s$-wave superconductor proximity coupled magnetic TIs is $\mathcal{H}_{\text{BdG}}=\sum_{\vec{k}}\Xi_{\vec{k}}^{\dag}H_{\text{BdG}}(\vec{k})\Xi_{\vec{k}}/2$, with $\Xi_{\vec{k}}=(\psi_{\vec{k}}^T,\psi_{-\vec{k}}^{\dag})^T$ and
\begin{equation}\label{BdG}
\begin{aligned}
H_{\text{BdG}}(\vec{k}) &=
\begin{pmatrix}
H_0(\vec{k})-\mu & \Delta_{\vec{k}}\\
\Delta^{\dag}_{\vec{k}} & -H_0^*(-\vec{k})+\mu
\end{pmatrix},
\end{aligned}
\end{equation}
Here $\mu$ is chemical potential, $\Delta_{\vec{k}}$ is the $s$-wave pairing function given by
\begin{equation}
\Delta_{\vec{k}}=
\begin{pmatrix}
i\Delta_1\sigma_2 & 0\\
0 & i\Delta_2\sigma_2
\end{pmatrix},
\end{equation}
where $\Delta_{1}$ ($\Delta_2$) is for top (bottom) surface.

\subsection{Phase diagram}
\label{phasediagram}

\begin{figure*}[t]
\begin{center}
\includegraphics[width=6.7in]{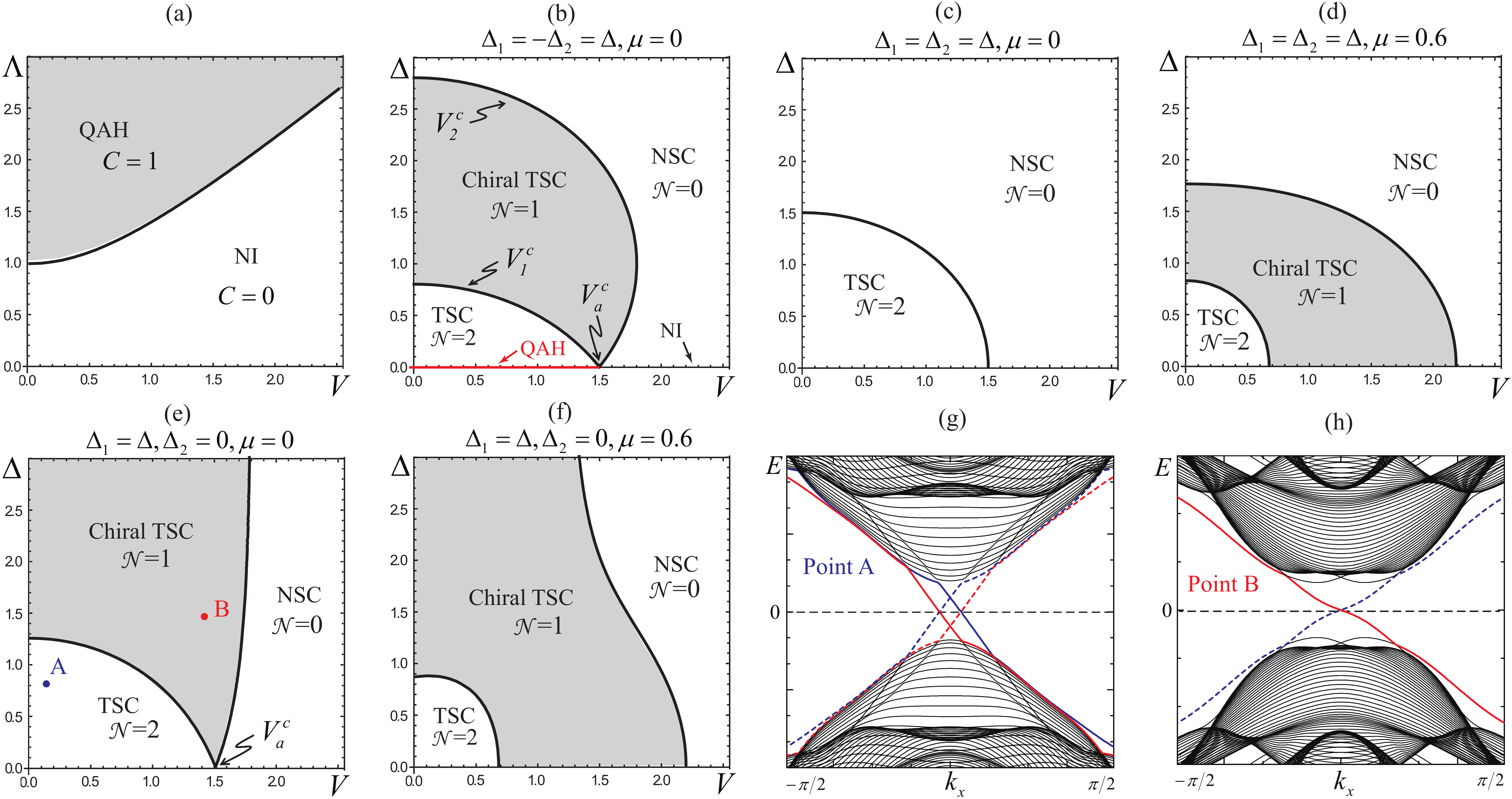}
\end{center}
\caption{(a) The phase diagram of the magnetic TI thin films for $m_0\neq0$ in the $(\Lambda, V)$ plane, only $\Lambda\geq0$ and $V\geq0$ is shown. (b)-(f) Phase diagram of the proximity coupled QAH system with typical parameters. Here $\Delta_1$, $\Delta_2$, $\mu$, $\Lambda$ are in the units of $|m_0|$, $m_1=1.0$, and $\Lambda=1.8$. In (b), phase QAH and NI are well defined only in the $\Delta=0$ line. (g) \& (h) Edge spectrum shows that chiral Majorana modes appears at each edge of the sample in the superconducting gap. The left (right) edge state are labeled by solid (dashed) line.}
\label{fig1}
\end{figure*}

First, we study the phase diagram of the system in Eq.~(\ref{BdG}) with $\mu$=0 and $\Delta_1$=$-\Delta_2$=$\Delta$. The bulk quasiparticle spectrum is
\begin{equation}
\begin{aligned}
E_{\kappa\eta\xi}(k) &= \kappa\Big(v_F^2k^2+V^2+\Lambda^2+(m(k)+\eta\Delta)^2
\\
&+2\xi
\sqrt{V^2v_F^2k^2+V^2\Lambda^2+\Lambda^2(m(k)+\eta\Delta)^2}\Big)^{\frac{1}{2}},
\end{aligned}
\end{equation}
with $\kappa,\eta,\xi=\pm$. Since the topological invariants only change with the bulk gap closes, we could find the phase boundaries by the gapless regions in the $(m,\Lambda,V,\Delta)$ space, and calculate the Chern number of each gapped phase. The parameters form a 4D space, and we are only interested in how the phases change as $V$ varies. Here we assume the parent magnetic TIs is already a QAH, with $\Lambda$ and $m_0$ fixed and $\Lambda>|m_0|$. The critical lines are determined by $(|m_0|\pm\Delta)^2+V^2=\Lambda^2$. The phase diagram is \emph{symmetric} with respect to $\Delta=0$ and $V=0$, so we only consider $\Delta>0$ and $V\geq0$ as shown in Fig.~\ref{fig1}(b). As expected, in the limit $\Delta=0$, the phase boundary reduces to the multicritical point $V^c_a$ between QAH and NI. For $V<\sqrt{\Lambda^2-(|m_0|+\Delta)^2}\equiv V_1^c$, the system is a nontrivial chiral TSC which is adiabatically connected to the QAH state in the $\Delta=0$ limit. The Chern number of such phase can be obtained in the $\Delta=0$ limit, in this case the BdG Hamiltonian in Eq.~(\ref{BdG}) is block diagonal. The total Chern number $\mathcal{N}=N_p+N_h$, where $N_p$ and $N_h$ are the Chern number of the particle and hole states, both of them are equal to that of $H_0(\vec{k})$, leading to $\mathcal{N}$=2. For $V>\sqrt{\Lambda^2-(|m_0|-\Delta)^2}\equiv V^c_{2}$, the system is adiabatically connected to a NI state so it must be a normal superconductor (NSC) phase with $\mathcal{N}$=0. For $V^c_1<V<V^c_2$, this state is adiabatically connected to the state in the limit of $V=0$ and $\Lambda-|m_0|<\Delta<\Lambda+|m_0|$~\cite{wang2015c}, the Chern number of which is $\mathcal{N}$=1. In other words, we have proven that SIA could drive the topological QPT and a chiral TSC phase with odd Chern number $\mathcal{N}$=1 is realized.

Next, we consider the phase diagram of Hamiltonian in Eq.~(\ref{BdG}) for more general values of $\Delta_1$ and $\Delta_2$. We assume $\Delta_2=\gamma\Delta_1\equiv\gamma\Delta$ and $\gamma$ is real. For $\mu=0$, the phase boundary is given by $\pm2\sqrt{\Lambda^2-V^2}\pm(1-\gamma)\Delta=\sqrt{4m_0^2+(1+\gamma)^2\Delta^2}$. Similar to the $\gamma=-1$ case, the Chern number of the gapped phases can be determined by its adiabatic connection to the $\Delta=0$ limit. For $\Delta_2=\Delta_1$, the $\mathcal{N}$=1 chiral TSC phase disappears due to accidental particle-hole symmetry in $H_0$ with $\mu=0$. As shown in Fig.~\ref{fig1}(e), when $\Delta_2$ reduces, $\mathcal{N}=1$ TSC phase emerges and is more favorable when $\gamma=0$, namely, only one surface has superconducting proximity effect. More generally, for the $\mu\neq0$ case, which corresponds to
the superconducting proximity effect of a doped QAH system, the proximity effect is effectively enhanced by the
finite density of states at the Fermi level. The phase space of the $\mathcal{N}$=1 chiral TSC enlarges from $\mu=0$ to $\mu\neq0$ as shown in Fig.~\ref{fig1}.

\subsection{Edge state}

To further confirm the system is exactly in the topological phase under such conditions, we study the edge spectrum in the tight-binding model which is obtained by regularizing the BdG Hamiltonian in Eq.~(\ref{BdG}) on a square lattice as $k_{x,y}\rightarrow a^{-1}\sin(k_{x,y}a)$ and $k_x^2+k_y^2=2-2a^{-2}(\cos(k_xa)+\cos(k_ya))$, where $a$ is the lattice constant. We consider $\mu=0$ and $\gamma=0$ pairing state in the cylindrical geometry with a periodic boundary condition in the $x$ direction and an open one in the $y$ direction. The energy spectrum for points $A$ and $B$ in Fig.~\ref{fig1}(e) are shown in Fig.~\ref{fig1}(g) and Fig.~\ref{fig1}(h), respectively. One can see that there are two chiral Majorana edge states for the $\mathcal{N}$=2 phase, and only one chiral Majorana state localized at opposite edges for the $\mathcal{N}$=1 phase, which are consistent with the previous study on bulk topological invariants.

\begin{table}[b]
\caption{The parameters of the 2D effective Hamiltonian in Eq.~(\ref{model0}) for 4 and 5 QL Cr$_{0.2}$(Bi$_{0.3}$Sb$_{0.6}$)$_2$Te$_3$ thin films. Each QL is about 1~nm thick. $\hbar$ is absorbed into $v_F$.}
\begin{center}\label{table1}
\renewcommand{\arraystretch}{1.5}
\begin{tabular*}{3.4in}
{@{\extracolsep{\fill}}cccccc}
\hline
\hline
& $v_F$ (eV \AA) & $m_0$ (eV) & $m_1$ (eV \AA$^2$) & $V$ (eV) & $\Lambda$ (eV)
\\
\hline
5 QL & $2.22$ & $+0.026$ & $13.7$ & $0.245\mathcal{E}$ & 0.048
\\
4 QL & $2.43$ & $-0.041$ & $30.2$ & $0.194\mathcal{E}$ & 0.045
\\
\hline
\hline
\end{tabular*}
\end{center}
\end{table}

\subsection{Estimation of SIA}

Now we have shown a stable $\mathcal{N}$=1 chiral TSC state can be induced by SIA from the QAH state in magnetic TIs. The advantage here by using an external electric field than a magnetic field is that the system is homogenous, where the magnetic domains of the material act as a single domain with up or down magnetization in the parent QAH state. Moreover, without invoking the external magnetic field at the coercivity, the superconducting proximity effect should be much stronger. To esitmate the magnitude of $V$ in realistic materials, we calculate the band structure of thin film Cr$_{0.2}$(Bi$_{0.3}$Sb$_{0.6}$)$_2$Te$_3$ in an external electric field along the $z$ direction. We consider the 3D bulk Hamiltonian $\mathcal{H}_{\mathrm{3D}}$~\cite{wang2013a} of magnetized TIs in a thin film configuration with thickness $d$, where the external electric field is modelled by adding $V_E(z)=\mathcal{E}z/d$. The parameters in $\mathcal{H}_{\mathrm{3D}}$ are taken from Ref.~\cite{zhang2009}, where the effect of reduced spin-orbit coupling strength resulting from the Cr substitution of (Bi, Sb) has been taken into account~\cite{zhang2013}.
The magnitude of the bulk exchange field is estimated as $\Lambda_b=yJ\langle S\rangle/2$, where $\langle S\rangle=3/2$ is the mean field value of local spin for Cr, the exchange coupling parameters between local spin and band electrons $J\approx2.7$~eV~\cite{yu2010}, and the magnetic dopants concentration $y=4\%$. We solve the eigen equation $[\mathcal{H}_{\mathrm{3D}}+V_E(z)]\psi_{n\vec{k}}(z)=E_{n\vec{k}}\psi_{n\vec{k}}(z)$
with open boundary condition $\psi_{n\vec{k}}(0)=\psi_{n\vec{k}}(d)=0$, where $n$ is the 2D subband index.
The parameters in Eq.~(\ref{model0}) for different quintuple layers (QL) can be obtained by projecting the bulk Hamiltonian onto the lowest four subbands, as shown in Table~\ref{table1}.

We can see that both 4~QL and 5~QL are QAH insulators in the presence of FM ordering. Without FM ordering, when $m_0m_1>0$, the system is NI; while $m_0m_1<0$, the system is QSH. The reduced $|m_0|$ and enhanced $\Lambda$ with increasing film thickness indicates larger gap and better QAH behavior in 5~QL than that in 4~QL~\cite{feng2016}. Oppositely, with finite superconducting pairing amplitude, the window of $V$ for $\mathcal{N}$=1 chiral TSC phase is narrower in 5~QL than that in 4~QL. For an estimation, taking $\Delta_2=0$, $\mu=0$, and $\Delta_1=3.0$~meV for superconductor Nb, the $\mathcal{N}$=1 chiral TSC emerges in 4~QL when $13.6~\text{meV}<V<20.9$~meV. With $0.25$~mm SrTiO$_3$ as the dielectric substrate, $12.9~\text{V}<V_{\text{bg}}<20.3$~V is needed to drive the system into $\mathcal{N}$=1 chiral TSC. Here $V_{\text{bg}}$ is the back-gate voltage on SrTiO$_3$. For comparison, with similar condition, the $\mathcal{N}$=1 chiral TSC window in 5~QL is $39.1~\text{meV}<V<41.1$~meV, and the corresponding back gate voltage is $29.9~\text{V}<V_{\text{bg}}<31.5$~V. The details on the estimation of $V_{\text{bg}}$ are in Appendix A.

\section{MBS}
\label{MBS}

\begin{figure}[b]
\begin{center}
\includegraphics[width=3.3in]{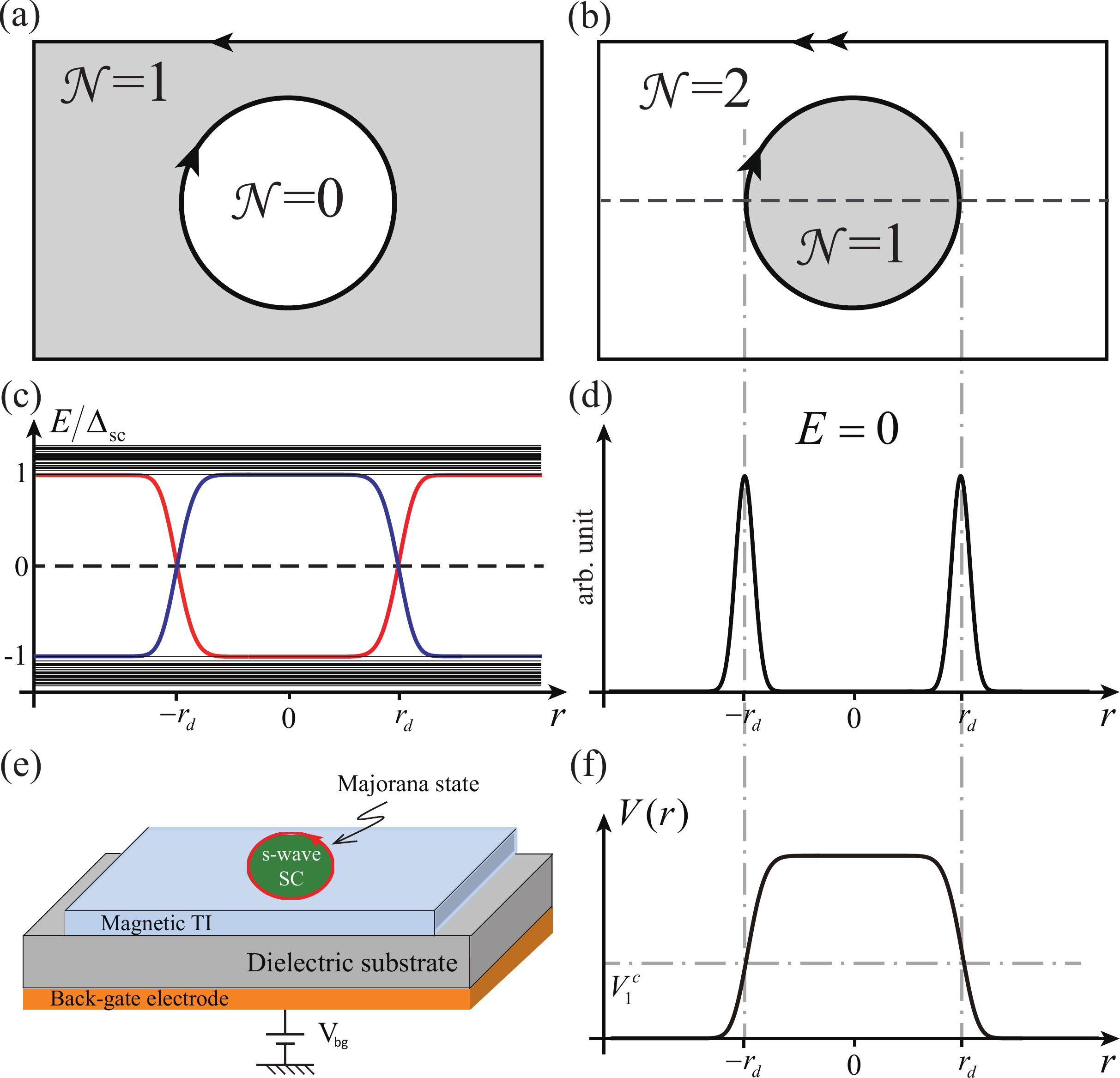}
\end{center}
\caption{(color online). Schematic illustration of a zero-energy MBS in the QAH-superconductor system. By applying a local electric field along the $z$ direction in the disk region, one may create (a) $\mathcal{N}$=0 spot in a $\mathcal{N}$=1 chiral TSC, or (b) $\mathcal{N}$=1 chiral TSC spot in a $\mathcal{N}$=2 TSC. The arrows represent Majorana edge modes. (c) $\rho(r,E)$. (d) Real-space probability distribution shows a zero-energy Majorana mode appears at the phase boundary along the radial direction. (e) Schematic of the device with $s$-wave superconductor island on top of the QAH in magnetic TI. (f) $V(r)$ with typical position dependence, $V(r)\propto\int^r_{-\infty}[e^{-(r'+r_d)^2}-e^{-(r'-r_d)^2}]dr'$.}
\label{fig2}
\end{figure}

The existence of a zero-energy MBS in the vortex core can be verified by using scanning tunneling microscopy (STM) and spectroscopy (STS) measurements of the local density of state (LDOS)~\cite{sun2016}. Here we predict the existence of a Majorana zero-energy state localized at the boundary of the applied electric field spot. We consider the disk region with radius $r_d$ in the system as shown in Fig.~\ref{fig2}. The spatial dependence of SIA term $V(x,y)$ can be controlled by the electric field through a local solid gate, where the inner region and outer region can have a different Chern number. Therefore, the gapless Majorana edge mode emerges at the phase boundary. These zero-energy edge modes are MBSs due to particle-hole symmetry, which can be obtained analytically by solving the BdG equation as in Appendix~\ref{Zero-energy MBS}. The explicit form of the zero-energy mode is
\begin{equation}
\varphi^{\eta\xi}_1(r)=\frac{N}{\sqrt{r}}\exp\left(\mp\frac{1}{v_F}\int^r_0E^0_{+\eta\xi}(r')dr'\right),
\end{equation}
where $N$ is the normalization factor, the finiteness of $\varphi_1(r)$ in the $r\rightarrow\infty$ limit determines the sign $\mp$. $E^0_{+\eta\xi}(r)$ are the eigenenergy of the BdG Hamiltonian in Eq.~(\ref{BdG}) when $\vec{k}=0$, and its spatial dependence comes from $V(r)$ or $\Delta(r)$. This zero-energy mode exists at the boundary where the sign of $E^0_{+\eta\xi}(r)$ changes. We assume $\Delta(r)$ is homogenous, and $V(r)$ is shown in Fig.~\ref{fig2}(f). For $r\gg r_d$, $V(r)=0$, while for $r\ll r_d$, $V_1^c<V(r)<V^c_2$, and $V(r_d)=V^c_1$. Therefore the gap closing point of $E^0_{+\eta\xi}(r)$ is $r=r_d$, which can be expanded as $E^0_{+\eta\xi}(r)=N_1(r-r_d)$. $N_1=(V(r_d)/\Lambda)\partial V(r_d)/\partial r$. Therefore the zero-energy edge mode is
\begin{equation}
\varphi^{\eta\xi}_1(r)=\frac{N'}{\sqrt{r}}\exp\left[-\frac{|N_1|}{2v_F}(r-r_d)^2\right],
\end{equation}
where $N'=N\exp(|N_1|r_d^2/2v_F)$. The edge mode is localized at $r=r_d$, with localization length $\lambda_M=\sqrt{v_F/|N_1|}$. In realistic systems, the electric field is not uniform. However, as we can see from the above calculation, the horizontal component of the electric field will not affect the MBS at edges. The vertical component of the inhomogeneous electric field determines the localization length of MBS. Furthermore if $r_d<\lambda_M$, the MBS will be gapped where the gap scales as $e^{-r_d/\lambda_M}$. For an estimation, for 4~QL magnetic TI, $\partial V(r_d)/\partial r\approx 4\times10^4$~eV/m, we get $\lambda_M\approx 141$~nm.

To identify TSC and observe the zero-energy Majorana state in this system, one can measure the LDOS by using STS with a superconducting tip. The spatial dispersion of the \emph{single} Majorana edge mode across the disk interface is characterized by a level crossing at $E=0$ and $r=r_d$ inside the superconducting gap $\Delta_{\text{sc}}$, as shown in Fig.~\ref{fig2}(c) the LDOS $\rho(r,E)$ at energy $E$ and position $r$ along the radial direction. The differential tunneling conductance $dI/dV$ maps in real space should see the edge modes on a single sharp circle with radius $r_d$, which is the zero-energy Majorana state. Such a single circle will evolve into two circles with different radius (one is larger than $r_d$, the other is smaller than $r_d$) at finite bias voltage, representing the energy evolution of the Majorana edge modes. This is the unique signature for the single Majorana edge state.

The applied electric field spot may be generated by a local back gate or a STM probe. For a local back gate, one may deposit the $s$-wave superconductor on top of the magnetic TI film; while for the STM probe, one may set the $s$-wave superconductor as a substrate. However, for either ways, it will be a great experimental challenge to fabricate such as a device and perform measurements. Practically, one can create an inhomogeneity in $\Delta(x,y)$ but with a homogenous $V$. This could also lead to chiral TSC phases and MBS at edges of $\Delta(x,y)$. The device is shown in Fig.~\ref{fig2}(e), an $s$-wave superconductor island may be grown on magnetic TI film with a global back gate. One may control topological QPT and observe the MBS at the boundary of $s$-wave superconductor island by tuning the back gate voltage.

\section{Helical TSC}
\label{helical}

\begin{figure}[t]
\begin{center}
\includegraphics[width=3.3in]{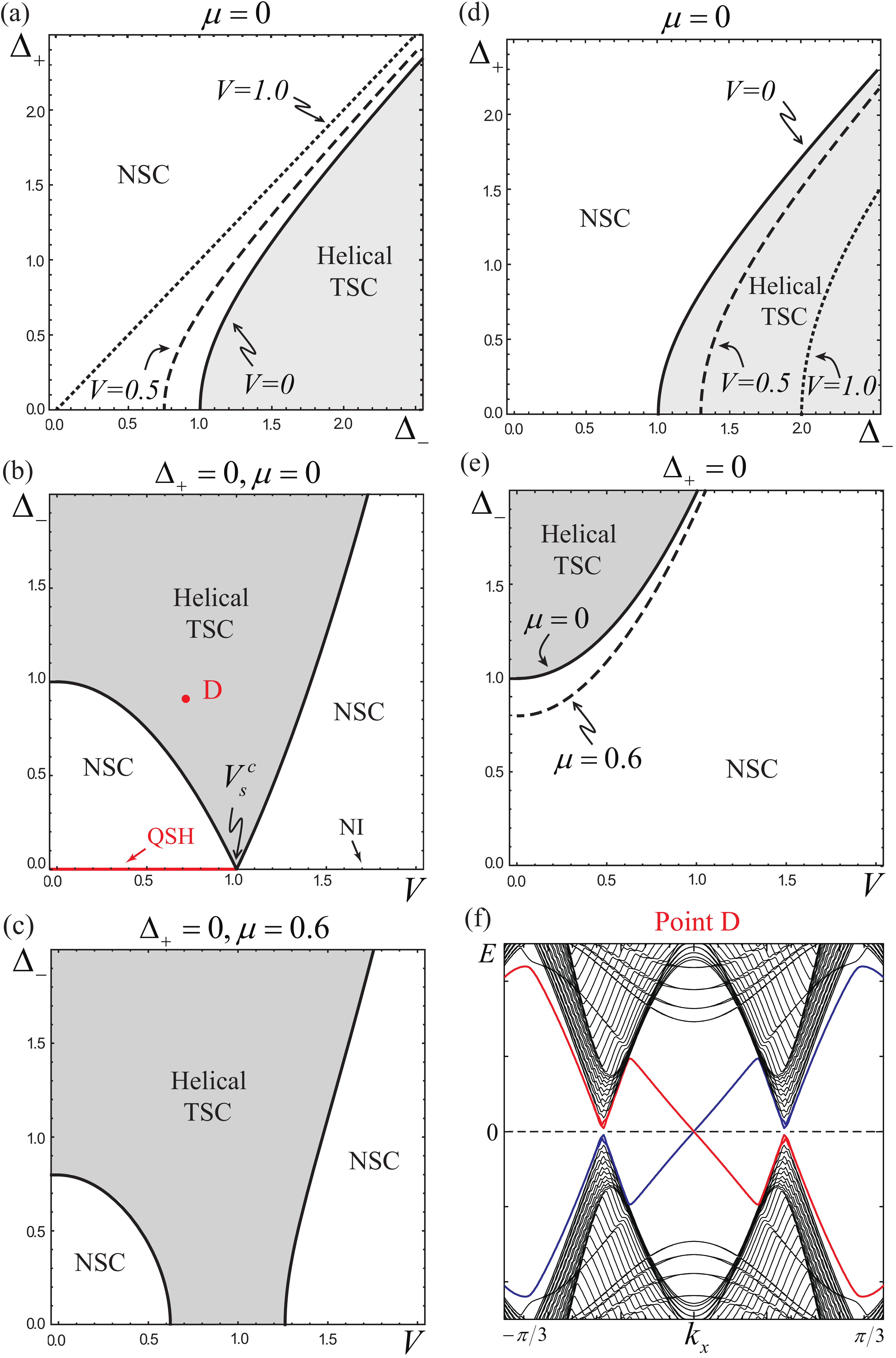}
\end{center}
\caption{(a)-(e) The phase diagram of the proximity coupled TI thin films for $m_0\neq0$ with several variables: $\Delta_+$, $\Delta_-$, $V$ and $\mu$; only $\Delta_{\pm}\geq0$ and $V\geq0$ is shown. Here $\Delta_+$, $\Delta_-$, $\mu$, $V$ are in the units of $|m_0|$. The parameters $v_F\equiv1$, $m_1\equiv1$; and $m_0=-1$ in (a)-(c), and $m_0=1$ in (d)-(e). In (b), phase QSH and NI are well defined only in the $\Delta_{\pm}=0$ line. (f) Edge spectrum shows a single helical edge state crossing the small bulk superconducting gap (at finite $k_x$) for point D.}
\label{fig3}
\end{figure}

Now we consider the TRI case. Intuitively, a QSH state with a single pair of helical edge states can be viewed as a $C$=1 QAH state together with its TR partner. As discussed in Sec.~\ref{chiral}, a QAH state with $C$=1 in proximity with an $s$-wave superconductor is a chiral TSC with $\mathcal{N}$=2. Therefore, with an infinitesimal $s$-wave superconductor proximity coupling, the QSH state can be naturally viewed as a superconductor with two pairs of helical Majorana edge states. However, according to the $Z_2$ classification in class DIII, such superconductor is topological trivial and even pairs of helical Majorana edge states are unstable under TRI perturbations~\cite{qi2009}. In other words, an infinitesimal superconducting gap drives the QSH phase into a TRI NSC with $Z_2$ index $\nu=0$. Quite different from the chiral TSC case, a helical TSC phase with $\nu=1$ does not necessarily emerges in the neighborhood of the transition between QSH and NI. Here we study under what condition the superconducting proximity coupled TI films would become a helical TSC. The TRI BdG Hamiltonian here is Eq.~(\ref{BdG}) with $\Lambda=0$, and $\Delta_{1(2)}$ is real. If $\mu=0$, the band dispersion is
\begin{equation}
\begin{aligned}
E_{\text{helical}}^{\kappa\eta\xi} &= \kappa\sqrt{(V+\eta v_Fk)^2+E_h^2},
\\
E_h =& \sqrt{m(k)^2+\Delta_+^2}+\xi|\Delta_-|,
\end{aligned}
\end{equation}
with $\kappa,\eta,\xi=\pm$, and $\Delta_{\pm}=(\Delta_1\pm\Delta_2)/2$. The phase boundary by the gapless regions is $V=\eta v_Fk$ and $m(k)^2+\Delta_+^2=\Delta^2_-$. The topological invariant of each gapped phase can be obtained in the $V=0$ limit. In this limit, when $m_0^2+\Delta_+^2>\Delta^2_-$, the system is adiabatically connected to a QSH or NI, so it must be a NSC. When $m_0^2+\Delta_+^2<\Delta^2_-$, the system should be a topologically nontrivial superconductor. The topological property of this state is clearly seen when $\Delta_+=0$, where a basis transformation decouples the BdG Hamiltonian into two models related by TR symmetry. Explicitly, one model is
\begin{equation}
H_1^{\text{BdG}}=v_F(k_y\sigma_1-k_x\sigma_2\varsigma_3)+m\sigma_3\varsigma_3-\Delta_-\sigma_2\varsigma_2,
\end{equation}
where $\varsigma_{2,3}$ are the Pauli matrices in Nambu space. The Chern number of $H_1^{\text{BdG}}$ is $\mathcal{N}$=1 when $|\Delta_-|>|m_0|$. Therefore, this state is a helical TSC~\cite{liu2011a} with a single pair of helical Majorana edge states. Fig.~\ref{fig3} shows the phase diagram with typical parameters, and several conclusions can be drawn: (i) helical TSC is maximized when $\Delta_+=0$, i.e., $\Delta_1$=$-\Delta_2$; (ii) if the parent system is a QSH ($m_0m_1<0$), the SIA will enlarge helical TSC in the phase diagram; while if the parent system is a NI ($m_0m_1>0$), finite SIA will always shrink helical TSC; (iii) helical TSC is more favorable with finite $\mu$ compared to $\mu=0$.

\section{Discussion}
\label{discussion}

The electrical control of TSC phases discussed above in TI thin films can be extended to van der Waals materials in 2D. The key point is that the orbitals which describes the low-energy physics are located on well-separated layers, therefore the band inversion and topological electronic properties can be controlled by an electric field. One interesting candidate is the 2D transition metal dichalcogenides \emph{MX}$_2$ class with 1T$'$ structure~\cite{qian2014}, where \emph{M} = (Mo, W) and \emph{X} = (Te, Se, S). The band inversion is between chalcogenide¡¯s $p$ and metal¡¯s $d$ orbitals, and it is tunable by a vertical electric field through dielectric layers. Together with the superconducting pairing and FM exchange coupling through the proximity effect, the chiral (helical) TSC state with an odd number of chiral (helical) Majorana edge states could be realized.

Recently, a new technique on gating device called ionic field-effect transistor has been developed~\cite{yu2015}. Compared to the conventional solid gate, tremendous electric field could be induced by the tunable Li ion intercalation. It is especially suitable for layered materials, and compatible with superconducting and FM proximity. The experimental progress on the gating method and material growth as well as rich material choice of TIs and van der Waals materials~\cite{novoselov2016} makes the realization of the 2D TSC states feasible.

\section{Conclusion}
\label{conclusion}

In summary, we propose the electric-field control of 2D TSC phases and Majorana fermions in TI films via the
proximity effect to an $s$-wave superconductor, which are generic for layered QAH and QSH insulator materials. For TRB case, we show that SIA could induce topological QPT and realize a $\mathcal{N}$=1 chiral TSC. Compared to the previous magnetic control, and electric manipulation should be more simpler and clearer. For TRI case, SIA could enlarge the helical TSC state in the phase diagram if the parent system is a QSH state. Moreover, instead of searching for large-gap QAH and QSH insulator materials for potential applications of the topologically protected conducting edge states, searching for small-gap QAH and QSH insulator materials as well as 2D Dirac and Weyl semimetals are even more useful for the realization of the TSC states in 2D. We hope the theoretical work here can aid the search for 2D TSC phases in hybrid systems.

\begin{acknowledgments}
The author is very much grateful to B.~Lian, Z.~Zhang, Y.~Wang, and S. C.~Zhang for valuable discussions. This work is supported by the National Thousand-Young-Talents Program; the National Key Research Program of China under Grant No.~2016YFA0300703; the Open Research Fund Program of the State Key Laboratory of Low-Dimensional Quantum Physics, through Contract No.~KF201606; and by Fudan University Initiative Scientific Research Program.
\end{acknowledgments}

\begin{appendix}

\section{Estimation of $V_{\mathrm{bg}}$}

We consider the device geometry where the insulating magnetic TI thin film is grown on the dielectric substrate such as SrTiO$_3$, and the top is covered with $s$-wave superconductor such as Nb. In this case, $\Delta_2=0$, $\mu=0$. Therefore, the window of $V$ for the $\mathcal{N}$=1 chiral TSC is between $2\sqrt{\Lambda^2-V^2}=\sqrt{4m_0^2+\Delta_1^2}\pm\Delta_1$. For an estimation, we take $\Delta_1=3.0$~meV for Nb film. From the numerical calculations, for 4~QL, the window of $V$ is $13.6~\text{meV}<V<20.9$~meV. Therefore, one can get the critical voltage window at the TI surfaces $V_{\mathrm{surf}}=\mathcal{E}/e$. From Table~\ref{table1}, $V=0.194\mathcal{E}$ for 4~QL. Thus, $0.070<V_{\mathrm{surf}}<0.108$~V. We use the capacitor model to estimate the value of the bottom gate voltage needed to drive such as transition into $\mathcal{N}$=1 chiral TSC,
\begin{equation}
\frac{\varepsilon_{\mathrm{TI}}V_{\mathrm{surf}}}{d_{\mathrm{TI}}}=\frac{\varepsilon_{\mathrm{b}}V_{\mathrm{bg}}}{d_{\mathrm{b}}},
\end{equation}
where $\varepsilon_{\mathrm{TI}}$ and $\varepsilon_{b}$ are dielectric constants for TI and dielectric substrate, respectively. $d_{\mathrm{TI}}$ and $d_{b}$ are thickness of TI and dielectric substrate. With $d_{\mathrm{TI}}=4$~nm, $d_{b}=0.25$~mm, and $\varepsilon_{\mathrm{TI}}\sim60$, dielectric substrate SrTiO$_3$ $\varepsilon_{b}\approx2\times10^4$. We get $12.9~\text{V}<V_{\text{bg}}<20.3$~V. Similar calculations can be done for 5~QL.

The voltage window for 5~QL to realize the $\mathcal{N}$=1 chiral TSC is narrower than that for 4~QL. The reason is that the QAH gap is larger in 5~QL, and to achieve chiral TSC, the superconducting proximity should exceed the QAH gap. Therefore, small-gap QAH insulator materials are more suitable for realizing the TSC states in 2D.

\section{Zero-energy MBS}
\label{Zero-energy MBS}

The zero-energy MBS at the boundary of the applied electric field spot with disk geometry can be obtained by analytically solving the BdG equation. For concreteness, we consider the BdG Hamiltonian in Eq.~(\ref{BdG}) with $\mu$=0 and $\Delta_1$=$-\Delta_2$=$\Delta$. The $8\times8$ model may be decomposed into four independent $2\times2$ Dirac Hamiltonians. $H_{\text{BdG}}(\vec{k})=H_{\text{BdG}}(0)+\delta H_{\text{BdG}}$, $H_{\text{BdG}}(0)$ can be diagonalized by a unitary transformation $U^\dag H_{\text{BdG}}(0)U=\text{diag}\{E^0_{\kappa\eta\xi}\}$, where $E^0_{\kappa\eta\xi}\equiv E_{\kappa\eta\xi}(0)$ and $\kappa,\eta,\xi=\pm$. Then the four $2\times2$ models are obtained by $U^\dag H_{\text{BdG}}(\vec{k})U$, and the explicit forms are,
\begin{equation}\label{BdG22}
H_D=\begin{pmatrix}
E^0_{+\eta\xi} & iv_{\eta\xi}k_-\\
-iv_{\eta\xi}k_+ & -E^0_{+\eta\xi}
\end{pmatrix}.
\end{equation}
Here to the lowest order, $v_{\eta\xi}=v_F$. For simplicity, in the following calculation we set $v_{\eta\xi}\equiv v_F$. Here $U=[(\phi_1,\phi_{2}),(\phi_3,\phi_{4}),(\phi_5,\phi_{6}),(\phi_7,\phi_{8})]$, and explicitly,
\begin{eqnarray}
\phi_{1,5} &&=\frac{1}{N_{1,5}}\begin{bmatrix}
0\\
\mp1\\
0\\
-\frac{\mp m+\Delta}{-V+\sqrt{(m\mp\Delta)^2+V^2}}\\
-\frac{-m\pm\Delta}{-V+\sqrt{(m\mp\Delta)^2+V^2}}\\
0\\
1\\
0
\end{bmatrix},
\\
\phi_{2,6} &&=\frac{1}{N_{2,6}}\begin{bmatrix}
\pm1\\
0\\
-\frac{\mp m+\Delta}{V+\sqrt{(m\mp\Delta)^2+V^2}}\\
0\\
0\\
-\frac{m\mp\Delta}{V+\sqrt{(m\mp\Delta)^2+V^2}}\\
0\\
1
\end{bmatrix},
\\
\phi_{3,7} &&=\frac{1}{N_{3,7}}\begin{bmatrix}
0\\
\mp1\\
0\\
-\frac{\pm m-\Delta}{V+\sqrt{(m\mp\Delta)^2+V^2}}\\
-\frac{m\mp\Delta}{V+\sqrt{(m\mp\Delta)^2+V^2}}\\
0\\
1\\
0
\end{bmatrix},
\\
\phi_{4,8} &&=\frac{1}{N_{4,8}}\begin{bmatrix}
\pm1\\
0\\
-\frac{\pm m-\Delta}{-V+\sqrt{(m\mp\Delta)^2+V^2}}\\
0\\
0\\
-\frac{-m\pm\Delta}{-V+\sqrt{(m\mp\Delta)^2+V^2}}\\
0\\
1
\end{bmatrix},
\end{eqnarray}
where $N_{i}$ ($i=1,2,\ldots,8$) are normalization factors.

The spatial dependence of $V(r)$ or $\Delta(r)$ will lead to the inhomogeneity of Dirac mass $E^0_{+\eta\xi}(r)$. In the disk geometry, we take the polar coordinate $(r,\theta)$. With $k_\pm=e^{\pm i\theta}[-i\partial/\partial r\pm(1/r)\partial/\partial\theta]$ and wavefunction ansatz $\Psi=[\varphi_1(r)e^{it\theta},\varphi_2(r)e^{i(t+1)\theta}]^T$, the radial and angular part of the Dirac Hamiltonian decouple as
\begin{equation}
H_D=\begin{pmatrix}
E^0_{+\eta\xi}(r) & v_F\Big(\frac{\partial}{\partial r}+\frac{(t+1)}{r}\Big)\\
v_F\Big(-\frac{\partial}{\partial r}+\frac{t}{r}\Big) & -E^0_{+\eta\xi}(r)
\end{pmatrix}.
\end{equation}
Since we are only interested in the zero-energy state, with ansatz $\varphi_2=\pm\varphi_1$, $H_D\Psi=0$ becomes
\begin{equation}
E^0_{+\eta\xi}(r)\varphi_1(r)\pm v_F\left(\frac{\partial}{\partial r}+\frac{1}{2r}\right)\varphi_1(r)=0.
\end{equation}
Therefore the solution is
\begin{equation}
\varphi^{\eta\xi}_1(r)=\frac{N}{\sqrt{r}}\exp\left(\mp\frac{1}{v_F}\int^r_0E^0_{+\eta\xi}(r')dr'\right),
\end{equation}
where $N$ is the normalization factor, the sign $\mp$ is determined by the finiteness of $\varphi_1(r)$ in the limit $r\rightarrow\infty$. This zero-energy mode exists at the boundary where the sign of $E^0_{+\eta\xi}(r)$ changes. The inhomogeneity in both $V(x,y)$ and $\Delta(x,y)$ will give rise to the gap closing in $E^0_{+\eta\xi}(r)$, thus lead to MBS.

\end{appendix}


\begin{thebibliography}{47}%
\makeatletter
\providecommand \@ifxundefined [1]{%
 \@ifx{#1\undefined}
}%
\providecommand \@ifnum [1]{%
 \ifnum #1\expandafter \@firstoftwo
 \else \expandafter \@secondoftwo
 \fi
}%
\providecommand \@ifx [1]{%
 \ifx #1\expandafter \@firstoftwo
 \else \expandafter \@secondoftwo
 \fi
}%
\providecommand \natexlab [1]{#1}%
\providecommand \enquote  [1]{``#1''}%
\providecommand \bibnamefont  [1]{#1}%
\providecommand \bibfnamefont [1]{#1}%
\providecommand \citenamefont [1]{#1}%
\providecommand \href@noop [0]{\@secondoftwo}%
\providecommand \href [0]{\begingroup \@sanitize@url \@href}%
\providecommand \@href[1]{\@@startlink{#1}\@@href}%
\providecommand \@@href[1]{\endgroup#1\@@endlink}%
\providecommand \@sanitize@url [0]{\catcode `\\12\catcode `\$12\catcode
  `\&12\catcode `\#12\catcode `\^12\catcode `\_12\catcode `\%12\relax}%
\providecommand \@@startlink[1]{}%
\providecommand \@@endlink[0]{}%
\providecommand \url  [0]{\begingroup\@sanitize@url \@url }%
\providecommand \@url [1]{\endgroup\@href {#1}{\urlprefix }}%
\providecommand \urlprefix  [0]{URL }%
\providecommand \Eprint [0]{\href }%
\providecommand \doibase [0]{http://dx.doi.org/}%
\providecommand \selectlanguage [0]{\@gobble}%
\providecommand \bibinfo  [0]{\@secondoftwo}%
\providecommand \bibfield  [0]{\@secondoftwo}%
\providecommand \translation [1]{[#1]}%
\providecommand \BibitemOpen [0]{}%
\providecommand \bibitemStop [0]{}%
\providecommand \bibitemNoStop [0]{.\EOS\space}%
\providecommand \EOS [0]{\spacefactor3000\relax}%
\providecommand \BibitemShut  [1]{\csname bibitem#1\endcsname}%
\let\auto@bib@innerbib\@empty
\bibitem [{\citenamefont {Hasan}\ and\ \citenamefont {Kane}(2010)}]{hasan2010}%
  \BibitemOpen
  \bibfield  {author} {\bibinfo {author} {\bibfnamefont {M.~Z.}\ \bibnamefont
  {Hasan}}\ and\ \bibinfo {author} {\bibfnamefont {C.~L.}\ \bibnamefont
  {Kane}},\ }\href {\doibase 10.1103/RevModPhys.82.3045} {\bibfield  {journal}
  {\bibinfo  {journal} {Rev. Mod. Phys.}\ }\textbf {\bibinfo {volume} {82}},\
  \bibinfo {pages} {3045} (\bibinfo {year} {2010})}\BibitemShut {NoStop}%
\bibitem [{\citenamefont {Qi}\ and\ \citenamefont {Zhang}(2011)}]{qi2011}%
  \BibitemOpen
  \bibfield  {author} {\bibinfo {author} {\bibfnamefont {X.-L.}\ \bibnamefont
  {Qi}}\ and\ \bibinfo {author} {\bibfnamefont {S.-C.}\ \bibnamefont {Zhang}},\
  }\href {\doibase 10.1103/RevModPhys.83.1057} {\bibfield  {journal} {\bibinfo
  {journal} {Rev. Mod. Phys.}\ }\textbf {\bibinfo {volume} {83}},\ \bibinfo
  {pages} {1057} (\bibinfo {year} {2011})}\BibitemShut {NoStop}%
\bibitem [{\citenamefont {Tanaka}\ \emph {et~al.}(2012)\citenamefont {Tanaka},
  \citenamefont {Sato},\ and\ \citenamefont {Nagaosa}}]{tanaka2012}%
  \BibitemOpen
  \bibfield  {author} {\bibinfo {author} {\bibfnamefont {Y.}~\bibnamefont
  {Tanaka}}, \bibinfo {author} {\bibfnamefont {M.}~\bibnamefont {Sato}}, \ and\
  \bibinfo {author} {\bibfnamefont {N.}~\bibnamefont {Nagaosa}},\ }\href
  {\doibase 10.1143/JPSJ.81.011013} {\bibfield  {journal} {\bibinfo  {journal}
  {J. Phys. Soc. Jpn.}\ }\textbf {\bibinfo {volume} {81}},\ \bibinfo {pages}
  {011013} (\bibinfo {year} {2012})}\BibitemShut {NoStop}%
\bibitem [{\citenamefont {Beenakker}\ and\ \citenamefont
  {Kouwenhoven}(2016)}]{beenakker2016}%
  \BibitemOpen
  \bibfield  {author} {\bibinfo {author} {\bibfnamefont {C.}~\bibnamefont
  {Beenakker}}\ and\ \bibinfo {author} {\bibfnamefont {L.}~\bibnamefont
  {Kouwenhoven}},\ }\href@noop {} {\bibfield  {journal} {\bibinfo  {journal}
  {Nature Phys.}\ }\textbf {\bibinfo {volume} {12}},\ \bibinfo {pages} {618}
  (\bibinfo {year} {2016})}\BibitemShut {NoStop}%
\bibitem [{\citenamefont {Sato}\ and\ \citenamefont {Ando}()}]{sato2016}%
  \BibitemOpen
  \bibfield  {author} {\bibinfo {author} {\bibfnamefont {M.}~\bibnamefont
  {Sato}}\ and\ \bibinfo {author} {\bibfnamefont {Y.}~\bibnamefont {Ando}},\
  }\href@noop {} {}\bibinfo {howpublished} {arXiv:1608.03395
  (2016)}\BibitemShut {NoStop}%
\bibitem [{\citenamefont {Nayak}\ \emph {et~al.}(2008)\citenamefont {Nayak},
  \citenamefont {Simon}, \citenamefont {Stern}, \citenamefont {Freedman},\ and\
  \citenamefont {Das~Sarma}}]{nayak2008}%
  \BibitemOpen
  \bibfield  {author} {\bibinfo {author} {\bibfnamefont {C.}~\bibnamefont
  {Nayak}}, \bibinfo {author} {\bibfnamefont {S.~H.}\ \bibnamefont {Simon}},
  \bibinfo {author} {\bibfnamefont {A.}~\bibnamefont {Stern}}, \bibinfo
  {author} {\bibfnamefont {M.}~\bibnamefont {Freedman}}, \ and\ \bibinfo
  {author} {\bibfnamefont {S.}~\bibnamefont {Das~Sarma}},\ }\href {\doibase
  10.1103/RevModPhys.80.1083} {\bibfield  {journal} {\bibinfo  {journal} {Rev.
  Mod. Phys.}\ }\textbf {\bibinfo {volume} {80}},\ \bibinfo {pages} {1083}
  (\bibinfo {year} {2008})}\BibitemShut {NoStop}%
\bibitem [{\citenamefont {Alicea}(2012)}]{alicea2012}%
  \BibitemOpen
  \bibfield  {author} {\bibinfo {author} {\bibfnamefont {J.}~\bibnamefont
  {Alicea}},\ }\href@noop {} {\bibfield  {journal} {\bibinfo  {journal} {Rep.
  Prog. Phys.}\ }\textbf {\bibinfo {volume} {75}},\ \bibinfo {pages} {076501}
  (\bibinfo {year} {2012})}\BibitemShut {NoStop}%
\bibitem [{\citenamefont {Beenakker}(2013)}]{beenakker2013}%
  \BibitemOpen
  \bibfield  {author} {\bibinfo {author} {\bibfnamefont {C.}~\bibnamefont
  {Beenakker}},\ }\href@noop {} {\bibfield  {journal} {\bibinfo  {journal}
  {Annu. Rev. Con. Mat. Phys.}\ }\textbf {\bibinfo {volume} {4}},\ \bibinfo
  {pages} {113} (\bibinfo {year} {2013})}\BibitemShut {NoStop}%
\bibitem [{\citenamefont {Schnyder}\ \emph {et~al.}(2008)\citenamefont
  {Schnyder}, \citenamefont {Ryu}, \citenamefont {Furusaki},\ and\
  \citenamefont {Ludwig}}]{schnyder2008}%
  \BibitemOpen
  \bibfield  {author} {\bibinfo {author} {\bibfnamefont {A.~P.}\ \bibnamefont
  {Schnyder}}, \bibinfo {author} {\bibfnamefont {S.}~\bibnamefont {Ryu}},
  \bibinfo {author} {\bibfnamefont {A.}~\bibnamefont {Furusaki}}, \ and\
  \bibinfo {author} {\bibfnamefont {A.~W.~W.}\ \bibnamefont {Ludwig}},\ }\href
  {\doibase 10.1103/PhysRevB.78.195125} {\bibfield  {journal} {\bibinfo
  {journal} {Phys. Rev. B}\ }\textbf {\bibinfo {volume} {78}},\ \bibinfo
  {pages} {195125} (\bibinfo {year} {2008})}\BibitemShut {NoStop}%
\bibitem [{\citenamefont {Read}\ and\ \citenamefont {Green}(2000)}]{read2000}%
  \BibitemOpen
  \bibfield  {author} {\bibinfo {author} {\bibfnamefont {N.}~\bibnamefont
  {Read}}\ and\ \bibinfo {author} {\bibfnamefont {D.}~\bibnamefont {Green}},\
  }\href {\doibase 10.1103/PhysRevB.61.10267} {\bibfield  {journal} {\bibinfo
  {journal} {Phys. Rev. B}\ }\textbf {\bibinfo {volume} {61}},\ \bibinfo
  {pages} {10267} (\bibinfo {year} {2000})}\BibitemShut {NoStop}%
\bibitem [{\citenamefont {Ivanov}(2001)}]{ivanov2001}%
  \BibitemOpen
  \bibfield  {author} {\bibinfo {author} {\bibfnamefont {D.~A.}\ \bibnamefont
  {Ivanov}},\ }\href {\doibase 10.1103/PhysRevLett.86.268} {\bibfield
  {journal} {\bibinfo  {journal} {Phys. Rev. Lett.}\ }\textbf {\bibinfo
  {volume} {86}},\ \bibinfo {pages} {268} (\bibinfo {year} {2001})}\BibitemShut
  {NoStop}%
\bibitem [{\citenamefont {Qi}\ \emph {et~al.}(2009)\citenamefont {Qi},
  \citenamefont {Hughes}, \citenamefont {Raghu},\ and\ \citenamefont
  {Zhang}}]{qi2009}%
  \BibitemOpen
  \bibfield  {author} {\bibinfo {author} {\bibfnamefont {X.-L.}\ \bibnamefont
  {Qi}}, \bibinfo {author} {\bibfnamefont {T.~L.}\ \bibnamefont {Hughes}},
  \bibinfo {author} {\bibfnamefont {S.}~\bibnamefont {Raghu}}, \ and\ \bibinfo
  {author} {\bibfnamefont {S.-C.}\ \bibnamefont {Zhang}},\ }\href {\doibase
  10.1103/PhysRevLett.102.187001} {\bibfield  {journal} {\bibinfo  {journal}
  {Phys. Rev. Lett.}\ }\textbf {\bibinfo {volume} {102}},\ \bibinfo {pages}
  {187001} (\bibinfo {year} {2009})}\BibitemShut {NoStop}%
\bibitem [{\citenamefont {Mackenzie}\ and\ \citenamefont
  {Maeno}(2003)}]{mackenzie2003}%
  \BibitemOpen
  \bibfield  {author} {\bibinfo {author} {\bibfnamefont {A.~P.}\ \bibnamefont
  {Mackenzie}}\ and\ \bibinfo {author} {\bibfnamefont {Y.}~\bibnamefont
  {Maeno}},\ }\href {\doibase 10.1103/RevModPhys.75.657} {\bibfield  {journal}
  {\bibinfo  {journal} {Rev. Mod. Phys.}\ }\textbf {\bibinfo {volume} {75}},\
  \bibinfo {pages} {657} (\bibinfo {year} {2003})}\BibitemShut {NoStop}%
\bibitem [{\citenamefont {Fu}\ and\ \citenamefont {Kane}(2009)}]{fu2009a}%
  \BibitemOpen
  \bibfield  {author} {\bibinfo {author} {\bibfnamefont {L.}~\bibnamefont
  {Fu}}\ and\ \bibinfo {author} {\bibfnamefont {C.~L.}\ \bibnamefont {Kane}},\
  }\href {\doibase 10.1103/PhysRevLett.102.216403} {\bibfield  {journal}
  {\bibinfo  {journal} {Phys. Rev. Lett.}\ }\textbf {\bibinfo {volume} {102}},\
  \bibinfo {pages} {216403} (\bibinfo {year} {2009})}\BibitemShut {NoStop}%
\bibitem [{\citenamefont {Sato}\ \emph {et~al.}(2009)\citenamefont {Sato},
  \citenamefont {Takahashi},\ and\ \citenamefont {Fujimoto}}]{sato2009}%
  \BibitemOpen
  \bibfield  {author} {\bibinfo {author} {\bibfnamefont {M.}~\bibnamefont
  {Sato}}, \bibinfo {author} {\bibfnamefont {Y.}~\bibnamefont {Takahashi}}, \
  and\ \bibinfo {author} {\bibfnamefont {S.}~\bibnamefont {Fujimoto}},\ }\href
  {\doibase 10.1103/PhysRevLett.103.020401} {\bibfield  {journal} {\bibinfo
  {journal} {Phys. Rev. Lett.}\ }\textbf {\bibinfo {volume} {103}},\ \bibinfo
  {pages} {020401} (\bibinfo {year} {2009})}\BibitemShut {NoStop}%
\bibitem [{\citenamefont {Sau}\ \emph {et~al.}(2010)\citenamefont {Sau},
  \citenamefont {Lutchyn}, \citenamefont {Tewari},\ and\ \citenamefont
  {Das~Sarma}}]{sau2010}%
  \BibitemOpen
  \bibfield  {author} {\bibinfo {author} {\bibfnamefont {J.~D.}\ \bibnamefont
  {Sau}}, \bibinfo {author} {\bibfnamefont {R.~M.}\ \bibnamefont {Lutchyn}},
  \bibinfo {author} {\bibfnamefont {S.}~\bibnamefont {Tewari}}, \ and\ \bibinfo
  {author} {\bibfnamefont {S.}~\bibnamefont {Das~Sarma}},\ }\href {\doibase
  10.1103/PhysRevLett.104.040502} {\bibfield  {journal} {\bibinfo  {journal}
  {Phys. Rev. Lett.}\ }\textbf {\bibinfo {volume} {104}},\ \bibinfo {pages}
  {040502} (\bibinfo {year} {2010})}\BibitemShut {NoStop}%
\bibitem [{\citenamefont {Alicea}(2010)}]{alicea2010}%
  \BibitemOpen
  \bibfield  {author} {\bibinfo {author} {\bibfnamefont {J.}~\bibnamefont
  {Alicea}},\ }\href {\doibase 10.1103/PhysRevB.81.125318} {\bibfield
  {journal} {\bibinfo  {journal} {Phys. Rev. B}\ }\textbf {\bibinfo {volume}
  {81}},\ \bibinfo {pages} {125318} (\bibinfo {year} {2010})}\BibitemShut
  {NoStop}%
\bibitem [{\citenamefont {Qi}\ \emph {et~al.}(2010)\citenamefont {Qi},
  \citenamefont {Hughes},\ and\ \citenamefont {Zhang}}]{qi2010b}%
  \BibitemOpen
  \bibfield  {author} {\bibinfo {author} {\bibfnamefont {X.-L.}\ \bibnamefont
  {Qi}}, \bibinfo {author} {\bibfnamefont {T.~L.}\ \bibnamefont {Hughes}}, \
  and\ \bibinfo {author} {\bibfnamefont {S.-C.}\ \bibnamefont {Zhang}},\ }\href
  {\doibase 10.1103/PhysRevB.82.184516} {\bibfield  {journal} {\bibinfo
  {journal} {Phys. Rev. B}\ }\textbf {\bibinfo {volume} {82}},\ \bibinfo
  {pages} {184516} (\bibinfo {year} {2010})}\BibitemShut {NoStop}%
\bibitem [{\citenamefont {Chung}\ \emph {et~al.}(2011)\citenamefont {Chung},
  \citenamefont {Qi}, \citenamefont {Maciejko},\ and\ \citenamefont
  {Zhang}}]{chung2011}%
  \BibitemOpen
  \bibfield  {author} {\bibinfo {author} {\bibfnamefont {S.~B.}\ \bibnamefont
  {Chung}}, \bibinfo {author} {\bibfnamefont {X.-L.}\ \bibnamefont {Qi}},
  \bibinfo {author} {\bibfnamefont {J.}~\bibnamefont {Maciejko}}, \ and\
  \bibinfo {author} {\bibfnamefont {S.-C.}\ \bibnamefont {Zhang}},\ }\href
  {\doibase 10.1103/PhysRevB.83.100512} {\bibfield  {journal} {\bibinfo
  {journal} {Phys. Rev. B}\ }\textbf {\bibinfo {volume} {83}},\ \bibinfo
  {pages} {100512} (\bibinfo {year} {2011})}\BibitemShut {NoStop}%
\bibitem [{\citenamefont {Wang}\ \emph
  {et~al.}(2015{\natexlab{a}})\citenamefont {Wang}, \citenamefont {Zhou},
  \citenamefont {Lian},\ and\ \citenamefont {Zhang}}]{wang2015c}%
  \BibitemOpen
  \bibfield  {author} {\bibinfo {author} {\bibfnamefont {J.}~\bibnamefont
  {Wang}}, \bibinfo {author} {\bibfnamefont {Q.}~\bibnamefont {Zhou}}, \bibinfo
  {author} {\bibfnamefont {B.}~\bibnamefont {Lian}}, \ and\ \bibinfo {author}
  {\bibfnamefont {S.-C.}\ \bibnamefont {Zhang}},\ }\href {\doibase
  10.1103/PhysRevB.92.064520} {\bibfield  {journal} {\bibinfo  {journal} {Phys.
  Rev. B}\ }\textbf {\bibinfo {volume} {92}},\ \bibinfo {pages} {064520}
  (\bibinfo {year} {2015}{\natexlab{a}})}\BibitemShut {NoStop}%
\bibitem [{\citenamefont {R\"ontynen}\ and\ \citenamefont
  {Ojanen}(2015)}]{ojanen2015}%
  \BibitemOpen
  \bibfield  {author} {\bibinfo {author} {\bibfnamefont {J.}~\bibnamefont
  {R\"ontynen}}\ and\ \bibinfo {author} {\bibfnamefont {T.}~\bibnamefont
  {Ojanen}},\ }\href {\doibase 10.1103/PhysRevLett.114.236803} {\bibfield
  {journal} {\bibinfo  {journal} {Phys. Rev. Lett.}\ }\textbf {\bibinfo
  {volume} {114}},\ \bibinfo {pages} {236803} (\bibinfo {year}
  {2015})}\BibitemShut {NoStop}%
\bibitem [{\citenamefont {Tanaka}\ \emph {et~al.}(2009)\citenamefont {Tanaka},
  \citenamefont {Yokoyama}, \citenamefont {Balatsky},\ and\ \citenamefont
  {Nagaosa}}]{tanaka2009}%
  \BibitemOpen
  \bibfield  {author} {\bibinfo {author} {\bibfnamefont {Y.}~\bibnamefont
  {Tanaka}}, \bibinfo {author} {\bibfnamefont {T.}~\bibnamefont {Yokoyama}},
  \bibinfo {author} {\bibfnamefont {A.~V.}\ \bibnamefont {Balatsky}}, \ and\
  \bibinfo {author} {\bibfnamefont {N.}~\bibnamefont {Nagaosa}},\ }\href
  {\doibase 10.1103/PhysRevB.79.060505} {\bibfield  {journal} {\bibinfo
  {journal} {Phys. Rev. B}\ }\textbf {\bibinfo {volume} {79}},\ \bibinfo
  {pages} {060505} (\bibinfo {year} {2009})}\BibitemShut {NoStop}%
\bibitem [{\citenamefont {Sato}\ and\ \citenamefont
  {Fujimoto}(2009)}]{sato2009a}%
  \BibitemOpen
  \bibfield  {author} {\bibinfo {author} {\bibfnamefont {M.}~\bibnamefont
  {Sato}}\ and\ \bibinfo {author} {\bibfnamefont {S.}~\bibnamefont
  {Fujimoto}},\ }\href {\doibase 10.1103/PhysRevB.79.094504} {\bibfield
  {journal} {\bibinfo  {journal} {Phys. Rev. B}\ }\textbf {\bibinfo {volume}
  {79}},\ \bibinfo {pages} {094504} (\bibinfo {year} {2009})}\BibitemShut
  {NoStop}%
\bibitem [{\citenamefont {Nakosai}\ \emph {et~al.}(2012)\citenamefont
  {Nakosai}, \citenamefont {Tanaka},\ and\ \citenamefont
  {Nagaosa}}]{nakosai2012}%
  \BibitemOpen
  \bibfield  {author} {\bibinfo {author} {\bibfnamefont {S.}~\bibnamefont
  {Nakosai}}, \bibinfo {author} {\bibfnamefont {Y.}~\bibnamefont {Tanaka}}, \
  and\ \bibinfo {author} {\bibfnamefont {N.}~\bibnamefont {Nagaosa}},\ }\href
  {\doibase 10.1103/PhysRevLett.108.147003} {\bibfield  {journal} {\bibinfo
  {journal} {Phys. Rev. Lett.}\ }\textbf {\bibinfo {volume} {108}},\ \bibinfo
  {pages} {147003} (\bibinfo {year} {2012})}\BibitemShut {NoStop}%
\bibitem [{\citenamefont {Wang}\ \emph
  {et~al.}(2014{\natexlab{a}})\citenamefont {Wang}, \citenamefont {Xu},\ and\
  \citenamefont {Zhang}}]{wang2014c}%
  \BibitemOpen
  \bibfield  {author} {\bibinfo {author} {\bibfnamefont {J.}~\bibnamefont
  {Wang}}, \bibinfo {author} {\bibfnamefont {Y.}~\bibnamefont {Xu}}, \ and\
  \bibinfo {author} {\bibfnamefont {S.-C.}\ \bibnamefont {Zhang}},\ }\href
  {\doibase 10.1103/PhysRevB.90.054503} {\bibfield  {journal} {\bibinfo
  {journal} {Phys. Rev. B}\ }\textbf {\bibinfo {volume} {90}},\ \bibinfo
  {pages} {054503} (\bibinfo {year} {2014}{\natexlab{a}})}\BibitemShut
  {NoStop}%
\bibitem [{\citenamefont {Chang}\ \emph {et~al.}(2013)\citenamefont {Chang},
  \citenamefont {Zhang}, \citenamefont {Feng}, \citenamefont {Shen},
  \citenamefont {Zhang}, \citenamefont {Guo}, \citenamefont {Li}, \citenamefont
  {Ou}, \citenamefont {Wei}, \citenamefont {Wang}, \citenamefont {Ji},
  \citenamefont {Feng}, \citenamefont {Ji}, \citenamefont {Chen}, \citenamefont
  {Jia}, \citenamefont {Dai}, \citenamefont {Fang}, \citenamefont {Zhang},
  \citenamefont {He}, \citenamefont {Wang}, \citenamefont {Lu}, \citenamefont
  {Ma},\ and\ \citenamefont {Xue}}]{chang2013b}%
  \BibitemOpen
  \bibfield  {author} {\bibinfo {author} {\bibfnamefont {C.-Z.}\ \bibnamefont
  {Chang}}, \bibinfo {author} {\bibfnamefont {J.}~\bibnamefont {Zhang}},
  \bibinfo {author} {\bibfnamefont {X.}~\bibnamefont {Feng}}, \bibinfo {author}
  {\bibfnamefont {J.}~\bibnamefont {Shen}}, \bibinfo {author} {\bibfnamefont
  {Z.}~\bibnamefont {Zhang}}, \bibinfo {author} {\bibfnamefont
  {M.}~\bibnamefont {Guo}}, \bibinfo {author} {\bibfnamefont {K.}~\bibnamefont
  {Li}}, \bibinfo {author} {\bibfnamefont {Y.}~\bibnamefont {Ou}}, \bibinfo
  {author} {\bibfnamefont {P.}~\bibnamefont {Wei}}, \bibinfo {author}
  {\bibfnamefont {L.-L.}\ \bibnamefont {Wang}}, \bibinfo {author}
  {\bibfnamefont {Z.-Q.}\ \bibnamefont {Ji}}, \bibinfo {author} {\bibfnamefont
  {Y.}~\bibnamefont {Feng}}, \bibinfo {author} {\bibfnamefont {S.}~\bibnamefont
  {Ji}}, \bibinfo {author} {\bibfnamefont {X.}~\bibnamefont {Chen}}, \bibinfo
  {author} {\bibfnamefont {J.}~\bibnamefont {Jia}}, \bibinfo {author}
  {\bibfnamefont {X.}~\bibnamefont {Dai}}, \bibinfo {author} {\bibfnamefont
  {Z.}~\bibnamefont {Fang}}, \bibinfo {author} {\bibfnamefont {S.-C.}\
  \bibnamefont {Zhang}}, \bibinfo {author} {\bibfnamefont {K.}~\bibnamefont
  {He}}, \bibinfo {author} {\bibfnamefont {Y.}~\bibnamefont {Wang}}, \bibinfo
  {author} {\bibfnamefont {L.}~\bibnamefont {Lu}}, \bibinfo {author}
  {\bibfnamefont {X.-C.}\ \bibnamefont {Ma}}, \ and\ \bibinfo {author}
  {\bibfnamefont {Q.-K.}\ \bibnamefont {Xue}},\ }\href {\doibase
  10.1126/science.1234414} {\bibfield  {journal} {\bibinfo  {journal}
  {Science}\ }\textbf {\bibinfo {volume} {340}},\ \bibinfo {pages} {167}
  (\bibinfo {year} {2013})}\BibitemShut {NoStop}%
\bibitem [{\citenamefont {Wang}\ \emph
  {et~al.}(2015{\natexlab{b}})\citenamefont {Wang}, \citenamefont {Lian},\ and\
  \citenamefont {Zhang}}]{wang2015d}%
  \BibitemOpen
  \bibfield  {author} {\bibinfo {author} {\bibfnamefont {J.}~\bibnamefont
  {Wang}}, \bibinfo {author} {\bibfnamefont {B.}~\bibnamefont {Lian}}, \ and\
  \bibinfo {author} {\bibfnamefont {S.-C.}\ \bibnamefont {Zhang}},\ }\href@noop
  {} {\bibfield  {journal} {\bibinfo  {journal} {Phys. Scr.}\ }\textbf
  {\bibinfo {volume} {T164}},\ \bibinfo {pages} {014003} (\bibinfo {year}
  {2015}{\natexlab{b}})}\BibitemShut {NoStop}%
\bibitem [{\citenamefont {He}\ \emph {et~al.}()\citenamefont {He},
  \citenamefont {Pan}, \citenamefont {Stern}, \citenamefont {Burks},
  \citenamefont {Che}, \citenamefont {Yin}, \citenamefont {Wang}, \citenamefont
  {Lian}, \citenamefont {Zhou}, \citenamefont {Choi}, \citenamefont {Murata},
  \citenamefont {Kou}, \citenamefont {Nie}, \citenamefont {Shao}, \citenamefont
  {Fan}, \citenamefont {Zhang}, \citenamefont {Liu}, \citenamefont {Xia},\ and\
  \citenamefont {Wang}}]{he2016}%
  \BibitemOpen
  \bibfield  {author} {\bibinfo {author} {\bibfnamefont {Q.~L.}\ \bibnamefont
  {He}}, \bibinfo {author} {\bibfnamefont {L.}~\bibnamefont {Pan}}, \bibinfo
  {author} {\bibfnamefont {A.~L.}\ \bibnamefont {Stern}}, \bibinfo {author}
  {\bibfnamefont {E.}~\bibnamefont {Burks}}, \bibinfo {author} {\bibfnamefont
  {X.}~\bibnamefont {Che}}, \bibinfo {author} {\bibfnamefont {G.}~\bibnamefont
  {Yin}}, \bibinfo {author} {\bibfnamefont {J.}~\bibnamefont {Wang}}, \bibinfo
  {author} {\bibfnamefont {B.}~\bibnamefont {Lian}}, \bibinfo {author}
  {\bibfnamefont {Q.}~\bibnamefont {Zhou}}, \bibinfo {author} {\bibfnamefont
  {E.~S.}\ \bibnamefont {Choi}}, \bibinfo {author} {\bibfnamefont
  {K.}~\bibnamefont {Murata}}, \bibinfo {author} {\bibfnamefont
  {X.}~\bibnamefont {Kou}}, \bibinfo {author} {\bibfnamefont {T.}~\bibnamefont
  {Nie}}, \bibinfo {author} {\bibfnamefont {Q.}~\bibnamefont {Shao}}, \bibinfo
  {author} {\bibfnamefont {Y.}~\bibnamefont {Fan}}, \bibinfo {author}
  {\bibfnamefont {S.-C.}\ \bibnamefont {Zhang}}, \bibinfo {author}
  {\bibfnamefont {K.}~\bibnamefont {Liu}}, \bibinfo {author} {\bibfnamefont
  {J.}~\bibnamefont {Xia}}, \ and\ \bibinfo {author} {\bibfnamefont {K.~L.}\
  \bibnamefont {Wang}},\ }\href@noop {} {}\bibinfo {howpublished} {arXiv:
  1606.05712 (2016)}\BibitemShut {NoStop}%
\bibitem [{\citenamefont {Lian}\ \emph {et~al.}(2016)\citenamefont {Lian},
  \citenamefont {Wang},\ and\ \citenamefont {Zhang}}]{lian2016}%
  \BibitemOpen
  \bibfield  {author} {\bibinfo {author} {\bibfnamefont {B.}~\bibnamefont
  {Lian}}, \bibinfo {author} {\bibfnamefont {J.}~\bibnamefont {Wang}}, \ and\
  \bibinfo {author} {\bibfnamefont {S.-C.}\ \bibnamefont {Zhang}},\ }\href
  {\doibase 10.1103/PhysRevB.93.161401} {\bibfield  {journal} {\bibinfo
  {journal} {Phys. Rev. B}\ }\textbf {\bibinfo {volume} {93}},\ \bibinfo
  {pages} {161401} (\bibinfo {year} {2016})}\BibitemShut {NoStop}%
\bibitem [{\citenamefont {Chen}\ \emph {et~al.}()\citenamefont {Chen},
  \citenamefont {He}, \citenamefont {Xu},\ and\ \citenamefont {Law}}]{law2016}%
  \BibitemOpen
  \bibfield  {author} {\bibinfo {author} {\bibfnamefont {C.-Z.}\ \bibnamefont
  {Chen}}, \bibinfo {author} {\bibfnamefont {J.~J.}\ \bibnamefont {He}},
  \bibinfo {author} {\bibfnamefont {D.-H.}\ \bibnamefont {Xu}}, \ and\ \bibinfo
  {author} {\bibfnamefont {K.~T.}\ \bibnamefont {Law}},\ }\href@noop {}
  {}\bibinfo {howpublished} {arXiv:1608.00237 (2016)}\BibitemShut {NoStop}%
\bibitem [{\citenamefont {Wang}\ \emph
  {et~al.}(2014{\natexlab{b}})\citenamefont {Wang}, \citenamefont {Lian},\ and\
  \citenamefont {Zhang}}]{wang2014a}%
  \BibitemOpen
  \bibfield  {author} {\bibinfo {author} {\bibfnamefont {J.}~\bibnamefont
  {Wang}}, \bibinfo {author} {\bibfnamefont {B.}~\bibnamefont {Lian}}, \ and\
  \bibinfo {author} {\bibfnamefont {S.-C.}\ \bibnamefont {Zhang}},\ }\href
  {\doibase 10.1103/PhysRevB.89.085106} {\bibfield  {journal} {\bibinfo
  {journal} {Phys. Rev. B}\ }\textbf {\bibinfo {volume} {89}},\ \bibinfo
  {pages} {085106} (\bibinfo {year} {2014}{\natexlab{b}})}\BibitemShut
  {NoStop}%
\bibitem [{\citenamefont {Liu}\ \emph {et~al.}(2016)\citenamefont {Liu},
  \citenamefont {Wang}, \citenamefont {Richardella}, \citenamefont {Kandala},
  \citenamefont {Li}, \citenamefont {Yazdani}, \citenamefont {Samarth},\ and\
  \citenamefont {Ong}}]{liu2016}%
  \BibitemOpen
  \bibfield  {author} {\bibinfo {author} {\bibfnamefont {M.}~\bibnamefont
  {Liu}}, \bibinfo {author} {\bibfnamefont {W.}~\bibnamefont {Wang}}, \bibinfo
  {author} {\bibfnamefont {A.~R.}\ \bibnamefont {Richardella}}, \bibinfo
  {author} {\bibfnamefont {A.}~\bibnamefont {Kandala}}, \bibinfo {author}
  {\bibfnamefont {J.}~\bibnamefont {Li}}, \bibinfo {author} {\bibfnamefont
  {A.}~\bibnamefont {Yazdani}}, \bibinfo {author} {\bibfnamefont
  {N.}~\bibnamefont {Samarth}}, \ and\ \bibinfo {author} {\bibfnamefont
  {N.~P.}\ \bibnamefont {Ong}},\ }\href@noop {} {\bibfield  {journal} {\bibinfo
   {journal} {Sci. Adv.}\ }\textbf {\bibinfo {volume} {2}},\ \bibinfo {pages}
  {e1600167} (\bibinfo {year} {2016})}\BibitemShut {NoStop}%
\bibitem [{\citenamefont {Chang}\ \emph {et~al.}(2015)\citenamefont {Chang},
  \citenamefont {Zhao}, \citenamefont {Kim}, \citenamefont {Zhang},
  \citenamefont {Assaf}, \citenamefont {Heiman}, \citenamefont {Zhang},
  \citenamefont {Liu}, \citenamefont {Chan},\ and\ \citenamefont
  {Moodera}}]{chang2015}%
  \BibitemOpen
  \bibfield  {author} {\bibinfo {author} {\bibfnamefont {C.-Z.}\ \bibnamefont
  {Chang}}, \bibinfo {author} {\bibfnamefont {W.}~\bibnamefont {Zhao}},
  \bibinfo {author} {\bibfnamefont {D.~Y.}\ \bibnamefont {Kim}}, \bibinfo
  {author} {\bibfnamefont {H.}~\bibnamefont {Zhang}}, \bibinfo {author}
  {\bibfnamefont {B.~A.}\ \bibnamefont {Assaf}}, \bibinfo {author}
  {\bibfnamefont {D.}~\bibnamefont {Heiman}}, \bibinfo {author} {\bibfnamefont
  {S.-C.}\ \bibnamefont {Zhang}}, \bibinfo {author} {\bibfnamefont
  {C.}~\bibnamefont {Liu}}, \bibinfo {author} {\bibfnamefont {M.~H.~W.}\
  \bibnamefont {Chan}}, \ and\ \bibinfo {author} {\bibfnamefont {J.~S.}\
  \bibnamefont {Moodera}},\ }\href@noop {} {\bibfield  {journal} {\bibinfo
  {journal} {Nature Mater.}\ }\textbf {\bibinfo {volume} {14}},\ \bibinfo
  {pages} {473} (\bibinfo {year} {2015})}\BibitemShut {NoStop}%
\bibitem [{\citenamefont {Kou}\ \emph {et~al.}(2015)\citenamefont {Kou},
  \citenamefont {Pan}, \citenamefont {Wang}, \citenamefont {Fan}, \citenamefont
  {Choi}, \citenamefont {Lee}, \citenamefont {Nie}, \citenamefont {Murata},
  \citenamefont {Shao}, \citenamefont {Zhang},\ and\ \citenamefont
  {Wang}}]{kou2015}%
  \BibitemOpen
  \bibfield  {author} {\bibinfo {author} {\bibfnamefont {X.}~\bibnamefont
  {Kou}}, \bibinfo {author} {\bibfnamefont {L.}~\bibnamefont {Pan}}, \bibinfo
  {author} {\bibfnamefont {J.}~\bibnamefont {Wang}}, \bibinfo {author}
  {\bibfnamefont {Y.}~\bibnamefont {Fan}}, \bibinfo {author} {\bibfnamefont
  {E.~S.}\ \bibnamefont {Choi}}, \bibinfo {author} {\bibfnamefont {W.-L.}\
  \bibnamefont {Lee}}, \bibinfo {author} {\bibfnamefont {T.}~\bibnamefont
  {Nie}}, \bibinfo {author} {\bibfnamefont {K.}~\bibnamefont {Murata}},
  \bibinfo {author} {\bibfnamefont {Q.}~\bibnamefont {Shao}}, \bibinfo {author}
  {\bibfnamefont {S.-C.}\ \bibnamefont {Zhang}}, \ and\ \bibinfo {author}
  {\bibfnamefont {K.~L.}\ \bibnamefont {Wang}},\ }\href@noop {} {\bibfield
  {journal} {\bibinfo  {journal} {Nat. Commun.}\ }\textbf {\bibinfo {volume}
  {6}},\ \bibinfo {pages} {8474} (\bibinfo {year} {2015})}\BibitemShut
  {NoStop}%
\bibitem [{\citenamefont {Feng}\ \emph {et~al.}(2015)\citenamefont {Feng},
  \citenamefont {Feng}, \citenamefont {Ou}, \citenamefont {Wang}, \citenamefont
  {Liu}, \citenamefont {Zhang}, \citenamefont {Zhao}, \citenamefont {Jiang},
  \citenamefont {Zhang}, \citenamefont {He}, \citenamefont {Ma}, \citenamefont
  {Xue},\ and\ \citenamefont {Wang}}]{fengy2015}%
  \BibitemOpen
  \bibfield  {author} {\bibinfo {author} {\bibfnamefont {Y.}~\bibnamefont
  {Feng}}, \bibinfo {author} {\bibfnamefont {X.}~\bibnamefont {Feng}}, \bibinfo
  {author} {\bibfnamefont {Y.}~\bibnamefont {Ou}}, \bibinfo {author}
  {\bibfnamefont {J.}~\bibnamefont {Wang}}, \bibinfo {author} {\bibfnamefont
  {C.}~\bibnamefont {Liu}}, \bibinfo {author} {\bibfnamefont {L.}~\bibnamefont
  {Zhang}}, \bibinfo {author} {\bibfnamefont {D.}~\bibnamefont {Zhao}},
  \bibinfo {author} {\bibfnamefont {G.}~\bibnamefont {Jiang}}, \bibinfo
  {author} {\bibfnamefont {S.-C.}\ \bibnamefont {Zhang}}, \bibinfo {author}
  {\bibfnamefont {K.}~\bibnamefont {He}}, \bibinfo {author} {\bibfnamefont
  {X.}~\bibnamefont {Ma}}, \bibinfo {author} {\bibfnamefont {Q.-K.}\
  \bibnamefont {Xue}}, \ and\ \bibinfo {author} {\bibfnamefont
  {Y.}~\bibnamefont {Wang}},\ }\href {\doibase 10.1103/PhysRevLett.115.126801}
  {\bibfield  {journal} {\bibinfo  {journal} {Phys. Rev. Lett.}\ }\textbf
  {\bibinfo {volume} {115}},\ \bibinfo {pages} {126801} (\bibinfo {year}
  {2015})}\BibitemShut {NoStop}%
\bibitem [{\citenamefont {Shan}\ \emph {et~al.}(2010)\citenamefont {Shan},
  \citenamefont {Lu},\ and\ \citenamefont {Shen}}]{shan2010}%
  \BibitemOpen
  \bibfield  {author} {\bibinfo {author} {\bibfnamefont {W.-Y.}\ \bibnamefont
  {Shan}}, \bibinfo {author} {\bibfnamefont {H.-Z.}\ \bibnamefont {Lu}}, \ and\
  \bibinfo {author} {\bibfnamefont {S.-Q.}\ \bibnamefont {Shen}},\ }\href@noop
  {} {\bibfield  {journal} {\bibinfo  {journal} {New J. Phys.}\ }\textbf
  {\bibinfo {volume} {12}},\ \bibinfo {pages} {043048} (\bibinfo {year}
  {2010})}\BibitemShut {NoStop}%
\bibitem [{\citenamefont {Wang}\ \emph
  {et~al.}(2015{\natexlab{c}})\citenamefont {Wang}, \citenamefont {Lian},\ and\
  \citenamefont {Zhang}}]{wang2015a}%
  \BibitemOpen
  \bibfield  {author} {\bibinfo {author} {\bibfnamefont {J.}~\bibnamefont
  {Wang}}, \bibinfo {author} {\bibfnamefont {B.}~\bibnamefont {Lian}}, \ and\
  \bibinfo {author} {\bibfnamefont {S.-C.}\ \bibnamefont {Zhang}},\ }\href
  {\doibase 10.1103/PhysRevLett.115.036805} {\bibfield  {journal} {\bibinfo
  {journal} {Phys. Rev. Lett.}\ }\textbf {\bibinfo {volume} {115}},\ \bibinfo
  {pages} {036805} (\bibinfo {year} {2015}{\natexlab{c}})}\BibitemShut
  {NoStop}%
\bibitem [{\citenamefont {Wang}\ \emph {et~al.}(2013)\citenamefont {Wang},
  \citenamefont {Lian}, \citenamefont {Zhang}, \citenamefont {Xu},\ and\
  \citenamefont {Zhang}}]{wang2013a}%
  \BibitemOpen
  \bibfield  {author} {\bibinfo {author} {\bibfnamefont {J.}~\bibnamefont
  {Wang}}, \bibinfo {author} {\bibfnamefont {B.}~\bibnamefont {Lian}}, \bibinfo
  {author} {\bibfnamefont {H.}~\bibnamefont {Zhang}}, \bibinfo {author}
  {\bibfnamefont {Y.}~\bibnamefont {Xu}}, \ and\ \bibinfo {author}
  {\bibfnamefont {S.-C.}\ \bibnamefont {Zhang}},\ }\href {\doibase
  10.1103/PhysRevLett.111.136801} {\bibfield  {journal} {\bibinfo  {journal}
  {Phys. Rev. Lett.}\ }\textbf {\bibinfo {volume} {111}},\ \bibinfo {pages}
  {136801} (\bibinfo {year} {2013})}\BibitemShut {NoStop}%
\bibitem [{\citenamefont {Zhang}\ \emph {et~al.}(2009)\citenamefont {Zhang},
  \citenamefont {Liu}, \citenamefont {Qi}, \citenamefont {Dai}, \citenamefont
  {Fang},\ and\ \citenamefont {Zhang}}]{zhang2009}%
  \BibitemOpen
  \bibfield  {author} {\bibinfo {author} {\bibfnamefont {H.}~\bibnamefont
  {Zhang}}, \bibinfo {author} {\bibfnamefont {C.-X.}\ \bibnamefont {Liu}},
  \bibinfo {author} {\bibfnamefont {X.-L.}\ \bibnamefont {Qi}}, \bibinfo
  {author} {\bibfnamefont {X.}~\bibnamefont {Dai}}, \bibinfo {author}
  {\bibfnamefont {Z.}~\bibnamefont {Fang}}, \ and\ \bibinfo {author}
  {\bibfnamefont {S.-C.}\ \bibnamefont {Zhang}},\ }\href@noop {} {\bibfield
  {journal} {\bibinfo  {journal} {Nature Phys.}\ }\textbf {\bibinfo {volume}
  {5}},\ \bibinfo {pages} {438} (\bibinfo {year} {2009})}\BibitemShut {NoStop}%
\bibitem [{\citenamefont {Zhang}\ \emph {et~al.}(2013)\citenamefont {Zhang},
  \citenamefont {Chang}, \citenamefont {Tang}, \citenamefont {Zhang},
  \citenamefont {Feng}, \citenamefont {Li}, \citenamefont {Wang}, \citenamefont
  {Chen}, \citenamefont {Liu}, \citenamefont {Duan}, \citenamefont {He},
  \citenamefont {Xue}, \citenamefont {Ma},\ and\ \citenamefont
  {Wang}}]{zhang2013}%
  \BibitemOpen
  \bibfield  {author} {\bibinfo {author} {\bibfnamefont {J.}~\bibnamefont
  {Zhang}}, \bibinfo {author} {\bibfnamefont {C.-Z.}\ \bibnamefont {Chang}},
  \bibinfo {author} {\bibfnamefont {P.}~\bibnamefont {Tang}}, \bibinfo {author}
  {\bibfnamefont {Z.}~\bibnamefont {Zhang}}, \bibinfo {author} {\bibfnamefont
  {X.}~\bibnamefont {Feng}}, \bibinfo {author} {\bibfnamefont {K.}~\bibnamefont
  {Li}}, \bibinfo {author} {\bibfnamefont {L.-l.}\ \bibnamefont {Wang}},
  \bibinfo {author} {\bibfnamefont {X.}~\bibnamefont {Chen}}, \bibinfo {author}
  {\bibfnamefont {C.}~\bibnamefont {Liu}}, \bibinfo {author} {\bibfnamefont
  {W.}~\bibnamefont {Duan}}, \bibinfo {author} {\bibfnamefont {K.}~\bibnamefont
  {He}}, \bibinfo {author} {\bibfnamefont {Q.-K.}\ \bibnamefont {Xue}},
  \bibinfo {author} {\bibfnamefont {X.}~\bibnamefont {Ma}}, \ and\ \bibinfo
  {author} {\bibfnamefont {Y.}~\bibnamefont {Wang}},\ }\href {\doibase
  10.1126/science.1230905} {\bibfield  {journal} {\bibinfo  {journal}
  {Science}\ }\textbf {\bibinfo {volume} {339}},\ \bibinfo {pages} {1582}
  (\bibinfo {year} {2013})}\BibitemShut {NoStop}%
\bibitem [{\citenamefont {Yu}\ \emph {et~al.}(2010)\citenamefont {Yu},
  \citenamefont {Zhang}, \citenamefont {Zhang}, \citenamefont {Zhang},
  \citenamefont {Dai},\ and\ \citenamefont {Fang}}]{yu2010}%
  \BibitemOpen
  \bibfield  {author} {\bibinfo {author} {\bibfnamefont {R.}~\bibnamefont
  {Yu}}, \bibinfo {author} {\bibfnamefont {W.}~\bibnamefont {Zhang}}, \bibinfo
  {author} {\bibfnamefont {H.-J.}\ \bibnamefont {Zhang}}, \bibinfo {author}
  {\bibfnamefont {S.-C.}\ \bibnamefont {Zhang}}, \bibinfo {author}
  {\bibfnamefont {X.}~\bibnamefont {Dai}}, \ and\ \bibinfo {author}
  {\bibfnamefont {Z.}~\bibnamefont {Fang}},\ }\href {\doibase
  10.1126/science.1187485} {\bibfield  {journal} {\bibinfo  {journal}
  {Science}\ }\textbf {\bibinfo {volume} {329}},\ \bibinfo {pages} {61}
  (\bibinfo {year} {2010})}\BibitemShut {NoStop}%
\bibitem [{\citenamefont {Feng}\ \emph {et~al.}(2016)\citenamefont {Feng},
  \citenamefont {Feng}, \citenamefont {Wang}, \citenamefont {Ou}, \citenamefont
  {Hao}, \citenamefont {Liu}, \citenamefont {Zhang}, \citenamefont {Zhang},
  \citenamefont {Lin}, \citenamefont {Liao}, \citenamefont {Li}, \citenamefont
  {Wang}, \citenamefont {Ji}, \citenamefont {Chen}, \citenamefont {Ma},
  \citenamefont {Zhang}, \citenamefont {Wang}, \citenamefont {He},\ and\
  \citenamefont {Xue}}]{feng2016}%
  \BibitemOpen
  \bibfield  {author} {\bibinfo {author} {\bibfnamefont {X.}~\bibnamefont
  {Feng}}, \bibinfo {author} {\bibfnamefont {Y.}~\bibnamefont {Feng}}, \bibinfo
  {author} {\bibfnamefont {J.}~\bibnamefont {Wang}}, \bibinfo {author}
  {\bibfnamefont {Y.}~\bibnamefont {Ou}}, \bibinfo {author} {\bibfnamefont
  {Z.}~\bibnamefont {Hao}}, \bibinfo {author} {\bibfnamefont {C.}~\bibnamefont
  {Liu}}, \bibinfo {author} {\bibfnamefont {Z.}~\bibnamefont {Zhang}}, \bibinfo
  {author} {\bibfnamefont {L.}~\bibnamefont {Zhang}}, \bibinfo {author}
  {\bibfnamefont {C.}~\bibnamefont {Lin}}, \bibinfo {author} {\bibfnamefont
  {J.}~\bibnamefont {Liao}}, \bibinfo {author} {\bibfnamefont {Y.}~\bibnamefont
  {Li}}, \bibinfo {author} {\bibfnamefont {L.-L.}\ \bibnamefont {Wang}},
  \bibinfo {author} {\bibfnamefont {S.-H.}\ \bibnamefont {Ji}}, \bibinfo
  {author} {\bibfnamefont {X.}~\bibnamefont {Chen}}, \bibinfo {author}
  {\bibfnamefont {X.}~\bibnamefont {Ma}}, \bibinfo {author} {\bibfnamefont
  {S.-C.}\ \bibnamefont {Zhang}}, \bibinfo {author} {\bibfnamefont
  {Y.}~\bibnamefont {Wang}}, \bibinfo {author} {\bibfnamefont {K.}~\bibnamefont
  {He}}, \ and\ \bibinfo {author} {\bibfnamefont {Q.-K.}\ \bibnamefont {Xue}},\
  }\href {\doibase 10.1002/adma.201600919} {\bibfield  {journal} {\bibinfo
  {journal} {Adv. Mat.}\ }\textbf {\bibinfo {volume} {28}},\ \bibinfo {pages}
  {6386} (\bibinfo {year} {2016})}\BibitemShut {NoStop}%
\bibitem [{\citenamefont {Sun}\ \emph {et~al.}(2016)\citenamefont {Sun},
  \citenamefont {Zhang}, \citenamefont {Hu}, \citenamefont {Li}, \citenamefont
  {Wang}, \citenamefont {Ma}, \citenamefont {Xu}, \citenamefont {Gao},
  \citenamefont {Guan}, \citenamefont {Li}, \citenamefont {Liu}, \citenamefont
  {Qian}, \citenamefont {Zhou}, \citenamefont {Fu}, \citenamefont {Li},
  \citenamefont {Zhang},\ and\ \citenamefont {Jia}}]{sun2016}%
  \BibitemOpen
  \bibfield  {author} {\bibinfo {author} {\bibfnamefont {H.-H.}\ \bibnamefont
  {Sun}}, \bibinfo {author} {\bibfnamefont {K.-W.}\ \bibnamefont {Zhang}},
  \bibinfo {author} {\bibfnamefont {L.-H.}\ \bibnamefont {Hu}}, \bibinfo
  {author} {\bibfnamefont {C.}~\bibnamefont {Li}}, \bibinfo {author}
  {\bibfnamefont {G.-Y.}\ \bibnamefont {Wang}}, \bibinfo {author}
  {\bibfnamefont {H.-Y.}\ \bibnamefont {Ma}}, \bibinfo {author} {\bibfnamefont
  {Z.-A.}\ \bibnamefont {Xu}}, \bibinfo {author} {\bibfnamefont {C.-L.}\
  \bibnamefont {Gao}}, \bibinfo {author} {\bibfnamefont {D.-D.}\ \bibnamefont
  {Guan}}, \bibinfo {author} {\bibfnamefont {Y.-Y.}\ \bibnamefont {Li}},
  \bibinfo {author} {\bibfnamefont {C.}~\bibnamefont {Liu}}, \bibinfo {author}
  {\bibfnamefont {D.}~\bibnamefont {Qian}}, \bibinfo {author} {\bibfnamefont
  {Y.}~\bibnamefont {Zhou}}, \bibinfo {author} {\bibfnamefont {L.}~\bibnamefont
  {Fu}}, \bibinfo {author} {\bibfnamefont {S.-C.}\ \bibnamefont {Li}}, \bibinfo
  {author} {\bibfnamefont {F.-C.}\ \bibnamefont {Zhang}}, \ and\ \bibinfo
  {author} {\bibfnamefont {J.-F.}\ \bibnamefont {Jia}},\ }\href {\doibase
  10.1103/PhysRevLett.116.257003} {\bibfield  {journal} {\bibinfo  {journal}
  {Phys. Rev. Lett.}\ }\textbf {\bibinfo {volume} {116}},\ \bibinfo {pages}
  {257003} (\bibinfo {year} {2016})}\BibitemShut {NoStop}%
\bibitem [{\citenamefont {Liu}\ and\ \citenamefont
  {Trauzettel}(2011)}]{liu2011a}%
  \BibitemOpen
  \bibfield  {author} {\bibinfo {author} {\bibfnamefont {C.-X.}\ \bibnamefont
  {Liu}}\ and\ \bibinfo {author} {\bibfnamefont {B.}~\bibnamefont
  {Trauzettel}},\ }\href {\doibase 10.1103/PhysRevB.83.220510} {\bibfield
  {journal} {\bibinfo  {journal} {Phys. Rev. B}\ }\textbf {\bibinfo {volume}
  {83}},\ \bibinfo {pages} {220510} (\bibinfo {year} {2011})}\BibitemShut
  {NoStop}%
\bibitem [{\citenamefont {Qian}\ \emph {et~al.}(2014)\citenamefont {Qian},
  \citenamefont {Liu}, \citenamefont {Fu},\ and\ \citenamefont
  {Li}}]{qian2014}%
  \BibitemOpen
  \bibfield  {author} {\bibinfo {author} {\bibfnamefont {X.}~\bibnamefont
  {Qian}}, \bibinfo {author} {\bibfnamefont {J.}~\bibnamefont {Liu}}, \bibinfo
  {author} {\bibfnamefont {L.}~\bibnamefont {Fu}}, \ and\ \bibinfo {author}
  {\bibfnamefont {J.}~\bibnamefont {Li}},\ }\href {\doibase
  10.1126/science.1256815} {\bibfield  {journal} {\bibinfo  {journal}
  {Science}\ }\textbf {\bibinfo {volume} {346}},\ \bibinfo {pages} {1344}
  (\bibinfo {year} {2014})}\BibitemShut {NoStop}%
\bibitem [{\citenamefont {Yu}\ \emph {et~al.}(2015)\citenamefont {Yu},
  \citenamefont {Yang}, \citenamefont {Lu}, \citenamefont {Yan}, \citenamefont
  {Cho}, \citenamefont {Ma}, \citenamefont {Niu}, \citenamefont {Kim},
  \citenamefont {Son}, \citenamefont {Feng}, \citenamefont {Li}, \citenamefont
  {Cheong}, \citenamefont {Chen},\ and\ \citenamefont {Zhang}}]{yu2015}%
  \BibitemOpen
  \bibfield  {author} {\bibinfo {author} {\bibfnamefont {Y.}~\bibnamefont
  {Yu}}, \bibinfo {author} {\bibfnamefont {F.}~\bibnamefont {Yang}}, \bibinfo
  {author} {\bibfnamefont {X.~F.}\ \bibnamefont {Lu}}, \bibinfo {author}
  {\bibfnamefont {Y.~J.}\ \bibnamefont {Yan}}, \bibinfo {author} {\bibfnamefont
  {Y.-H.}\ \bibnamefont {Cho}}, \bibinfo {author} {\bibfnamefont
  {L.}~\bibnamefont {Ma}}, \bibinfo {author} {\bibfnamefont {X.}~\bibnamefont
  {Niu}}, \bibinfo {author} {\bibfnamefont {S.}~\bibnamefont {Kim}}, \bibinfo
  {author} {\bibfnamefont {Y.-W.}\ \bibnamefont {Son}}, \bibinfo {author}
  {\bibfnamefont {D.}~\bibnamefont {Feng}}, \bibinfo {author} {\bibfnamefont
  {S.}~\bibnamefont {Li}}, \bibinfo {author} {\bibfnamefont {S.-W.}\
  \bibnamefont {Cheong}}, \bibinfo {author} {\bibfnamefont {X.~H.}\
  \bibnamefont {Chen}}, \ and\ \bibinfo {author} {\bibfnamefont
  {Y.}~\bibnamefont {Zhang}},\ }\href@noop {} {\bibfield  {journal} {\bibinfo
  {journal} {Nature Nano.}\ }\textbf {\bibinfo {volume} {10}},\ \bibinfo
  {pages} {270} (\bibinfo {year} {2015})}\BibitemShut {NoStop}%
\bibitem [{\citenamefont {Novoselov}\ \emph {et~al.}(2016)\citenamefont
  {Novoselov}, \citenamefont {Mishchenko}, \citenamefont {Carvalho},\ and\
  \citenamefont {Castro~Neto}}]{novoselov2016}%
  \BibitemOpen
  \bibfield  {author} {\bibinfo {author} {\bibfnamefont {K.~S.}\ \bibnamefont
  {Novoselov}}, \bibinfo {author} {\bibfnamefont {A.}~\bibnamefont
  {Mishchenko}}, \bibinfo {author} {\bibfnamefont {A.}~\bibnamefont
  {Carvalho}}, \ and\ \bibinfo {author} {\bibfnamefont {A.~H.}\ \bibnamefont
  {Castro~Neto}},\ }\href@noop {} {\bibfield  {journal} {\bibinfo  {journal}
  {Science}\ }\textbf {\bibinfo {volume} {353}},\ \bibinfo {pages} {aac9439}
  (\bibinfo {year} {2016})}\BibitemShut {NoStop}%
\end{thebibliography}
\end{document}